\documentclass[conference]{IEEEtran}
\makeatletter
\def\ps@headings{%
\def\@oddhead{\mbox{}\scriptsize\rightmark \hfil \thepage}%
\def\@evenhead{\scriptsize\thepage \hfil \leftmark\mbox{}}%
\def\@oddfoot{}%
\def\@evenfoot{}}
\makeatother
\usepackage{amsmath}
\usepackage{amssymb}
\usepackage{amsmath}
\usepackage{amssymb}
\usepackage{epsfig}
\usepackage{epsf}
\usepackage{subfigure}
\usepackage{graphicx}
\usepackage{url}
\usepackage[]{authblk}
\usepackage{cite}

\def\BibTeX{{\rm B\kern-.05em{\sc i\kern-.025em b}\kern-.08em
    T\kern-.1667em\lower.7ex\hbox{E}\kern-.125emX}}

\newtheorem{theorem}{Theorem}

\newtheorem{lemma}{Lemma}

\newcounter{numcount}

\newcommand{\cnt}{\Roman{numcount}\;\stepcounter{numcount}}

\begin{document}

\newcommand{\caseone} { { \nearrow } { \hspace{-3.5mm} \searrow } {\hspace{-3.55mm}  \raisebox{4.7pt}{{$\rightharpoonup$}}} {\hspace{-3.5mm}  \raisebox{-4.7pt}{{$\rightharpoondown$}}} }

\newcommand{\casetwo} { { \searrow } {\hspace{-3.5mm}  \raisebox{4.7pt}{{$\rightarrow$}}} {\hspace{-3.5mm}  \raisebox{-4.7pt}{{$\rightharpoondown$}}} }

\newcommand{\casethree} { { \nearrow } {\hspace{-3.5mm}  \raisebox{4.7pt}{{$\rightarrow$}}} {\hspace{-3.5mm}  \raisebox{-4.7pt}{{$\rightharpoondown$}}} }

\newcommand{\casefour} { { \raisebox{4.7pt}{{$\rightarrow$}}} {\hspace{-3.5mm}  \raisebox{-4.7pt}{{$\rightarrow$}}} }

\newcommand{\casefive} { { \raisebox{4.7pt}{{$\rightarrow$}}} }

\newcommand{\casesix} { { \searrow } {\hspace{-3.5mm}  \raisebox{4.7pt}{{$\rightarrow$}}}  }

\newcommand{\caseseven} { { \nearrow } {\hspace{-3.5mm}  \raisebox{4.7pt}{{$\rightharpoonup$}}} }

\newcommand{\caseeight} { { \nearrow } { \hspace{-3.5mm} \searrow } {\hspace{-3.5mm}  \raisebox{4.7pt}{{$\rightharpoonup$}}} }

\newcommand{\casefifteen} { { \nearrow } {\hspace{-3.5mm} \searrow } }

%%%%%%%%%%%%%%%%%%%%%%%%%%%%%%%%%%%%%%%%%%%%%%%%%%%
%%%%%%%%%%%%%%%%%%%%%%%%%%%%%%%%%%%%%%%%%%%%%%%%%%%
%%%%%%%%%%%%%%%%%%%%%%%%%%%%%%%%%%%%%%%%%%%%%%%%%%%

\title{Communication Through Collisions:\\
Opportunistic Utilization of Past Receptions}

\author{Alireza~Vahid$^\dagger$,
        Mohammad~Ali~Maddah-Ali$^\ast$,
        and~A.~Salman~Avestimehr$^\dagger$\\
				$^\dagger$Cornell University, Ithaca, NY, USA\\ 
				$^\ast$Bell Labs, Alcatel-Lucent, Holmdel, NJ, USA      				
}

%\author{Alireza~Vahid,
%        Mohammad~Ali~Maddah-Ali,
%        and~A.~Salman~Avestimehr
%        \thanks{A. Vahid and A. S. Avestimehr are with the School of Electrical and Computer Engineering, Cornell University, Ithaca, NY, USA. Email: {\sffamily av292@cornell.edu} and {\sffamily avestimehr@ece.cornell.edu}.}
%\thanks{Mohammad~Ali~Maddah-Ali is with Bell Labs, Alcatel-Lucent, Holmdel, NJ, USA. Email: {\sffamily mohammadali.maddah-ali@alcatel-lucent.com}.}
%\thanks{The work of A. S. Avestimehr and A. Vahid is in part supported by NSF Grants CAREER-0953117, CCF-1161720, NETS-1161904, AFOSR Young Investigator Program Award, and ONR award N000141310094.}
%}

% The distinguishing features of a wireless network compared to a wired network are broadcast and superposition, meaning that the signal transmitted by a node may reach several nodes, and a node may receive signals from several other nodes simultaneously. The unwanted signal at each receiver is treated either as interference or as collision. Regardless of how we treat the unwanted signal, it is considered harmful to communication rather than blessing. 

\maketitle
%%%%%%%%%%%%%%%%%%%%%%%%%%%%%%%%%%%%%%%%%%%%%%%%%%%
\begin{abstract}
When several wireless users are sharing the spectrum, packet collision is a simple, yet widely used model for interference. Under this model, when transmitters cause interference at any of the receivers, their collided packets are discarded and need to be retransmitted. However, in reality, that receiver can still store its analog received signal and utilize it for decoding the packets in the future (for example, by successive interference cancellation techniques). In this work, we propose a physical layer model for wireless packet networks that allows for such flexibility at the receivers. We assume that the transmitters will be aware of the state of the channel (\emph{i.e.} when and where collisions occur, or an unintended receiver overhears the signal) with some delay, and propose several coding opportunities that can be utilized by the transmitters to exploit the available signal at the receivers for interference management (as opposed to discarding them). We analyze the achievable throughput of our strategy in a canonical  interference channel with two transmitter-receiver pairs, and demonstrate the gain over conventional schemes. By deriving an outer-bound, we also prove the optimality of our scheme for the corresponding model.
\end{abstract}
%%%%%%%%%%%%%%%%%%%%%%%%%%%%%%%%%%%%%%%%%%%%%%%%%%%


\section{Introduction}
\label{Sec:Introduction}

The packet collision model is a simple, yet widely used model for interference in wireless packet networks. Under this model, when transmitters cause interference at any of the receivers, their collided packets are discarded and need to be retransmitted. As a result, centralized scheduling or Aloha-type mechanisms are used to minimize the impact of collisions. However, it is widely known that when collision occurs, the receiver can still store its analog received signal and utilize it for decoding the packets in the future. This can be done via a variety of techniques studied in multiuser joint detection and interference cancellation for cellular networks (\emph{e.g.},~\cite{boutros2002iterative,verdu1998multiuser,reynolds2002blind,el2000iterative,poor2004iterative}). ZigZag decoding~\cite{gollakota2008zigzag} also demonstrates that interference decoding and successive interference cancellation can be accomplished in 802.11 MAC. 

% In this paper, we propose a physical layer model for packet networks that allows the flexibility of storing the analog received signals at the receivers. The stored analog signals can be utilized for decoding packets in the future. 

\begin{figure}[h]
\centering
\includegraphics[height = 5.5cm]{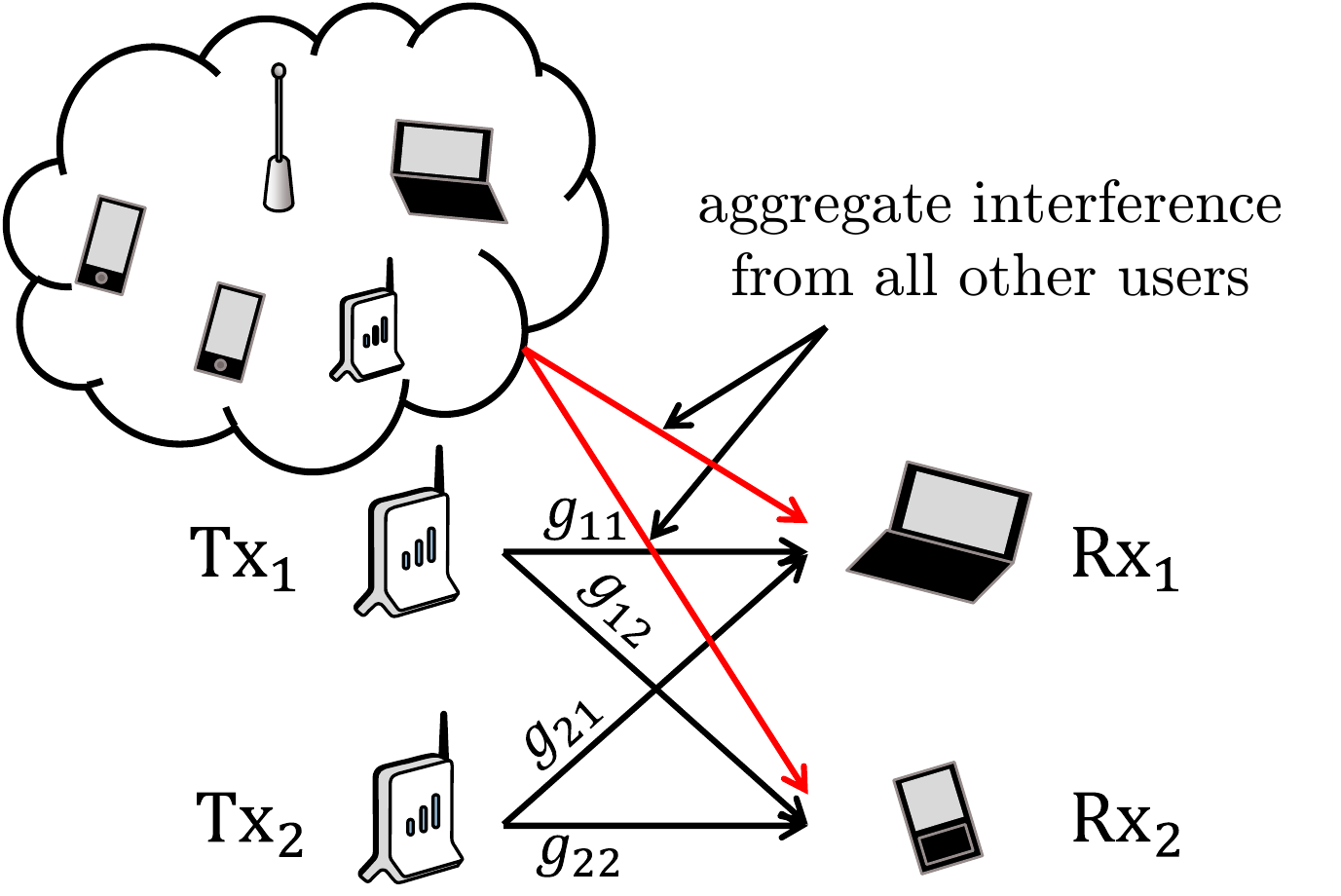}
\caption{A wireless packet network in which multiple transmitter-receiver pairs are communicating with each other. We focus on two nearby pairs that can coordinate for interference management.\label{Fig:WLAN}}
\end{figure}

In this paper, we propose a physical layer model for packet networks that allows the flexibility of storing the analog received signals at the receivers, so that they can be utilized for decoding packets in the future. We study the coding opportunities that arise for interference management in such networks. We focus on a network in which  two transmitter-receiver pairs can coordinate for interference management, based on their \emph{delayed} knowledge of the channel state, see Fig.~\ref{Fig:WLAN}\footnote{Note that the transmitters do not exchange any information bits and they solely coordinate via their delayed knowledge of the channel-state.}. Depending on the aggregate interference from other users, there will be four channel states at each receiver, say ${\sf Rx}_1$, as follows: State~1: if the signal-to-interference-plus-noise ratio (SINR) of the link from ${\sf Tx}_1$ is above a threshold, we assume that ${\sf Rx}_1$ can decode its intended packet; State~2: if the SINR of the link from ${\sf Tx}_2$ is above a threshold, we assume that ${\sf Rx}_1$ can decode other user's packet; State~3: if the SINR of both incoming links is below the desired threshold, but each link is individually strong, then we assume that ${\sf Rx}_1$ obtains a linear combination of the transmitted packets; finally, State~4: in any other scenario, ${\sf Rx}_1$ discards the received signal. Motivated by recent results in information theory on the gains obtained from completely outdated channel state information (e.g.,~\cite{Maddah-Tse-Allerton,Jafar_Retrospective,Abdoli2011IC-X-Arxiv,Vaze_DCSIT_MIMO_BC,BFICAllerton,SinaXChannel1,MultiHopDelayed}), we study how transmitters can optimally utilize their delayed knowledge for packet coding and interference management.

% We note that the transmitters do not exchange information and they solely coordinate via their delayed knowledge of the channel. 

Our main contributions in this paper are threefold. First, we develop two novel coding opportunities at the transmitters for interference management that go well beyond the conventional approach of packet repetition coding. Suppose each transmitter sends one packet and each receiver obtains a linear combination of the packets. Then, it will be sufficient to deliver only one of the packets to both receivers in the future. We refer to such packets as ``packets of common interest''. In the first coding opportunity, ``packet delivery with side information'', at each transmitter, we combine a packet of common interest and a packet that was overheard by the unintended receiver. We show that delivering such combination of packets can resolve two previous interferences. In the second coding opportunity, ``interference delivery with side information'', we take advantage of the overheard packets and the packets that are available at their corresponding receivers but are still useful for the unintended receiver. In fact, we combine packets that were transmitted in three different cases, and we show that by delivering such combination of packets, we can resolve interference in those three cases.

Second, we propose a strategy that systematically utilizes the aforementioned coding opportunities in a network in which two transmitter-receiver pairs can coordinate for interference management. Our transmission strategy is carried on over two phases. Each channel realization creates coding opportunities that can be exploited in the second phase. After the initial phase, we update the status of the previously transmitted packets by moving them to a number of virtual queues. Then, we incorporate our coding ideas to empty these queues at higher throughput. We observe that \emph{merging} or \emph{concatenating} some of the opportunities can offer even more gain. To achieve the optimal throughput, we find the most efficient arrangement of combination, concatenation, and merging of the opportunities. 

Third, we show the optimality of our transmission strategy, and we characterize the throughput region of a network in which two transmitter-receiver pairs can coordinate for interference management. We derive an information theoretic outer-bound for our problem. The key idea for the derivation of the outer-bound is an extremal rank inequality for an underlying broadcast channel, which leads to a tight outer-bound for our problem. The established inequality provides a bound on how much a transmitter can favor one receiver to the other in terms of the rank of the received signal at the two receivers, using delayed knowledge of the channel state information. Our achievable throughput region agrees with the information theoretic outer-bound we obtain, thus proving the optimality of our scheme.

The rest of the paper is organized as follows. In Section~\ref{Sec:System}, we formulate our problem and present our main results. We then provide an overview of our main achievability techniques in Section~\ref{Sec:Opportunities}. In Section~\ref{Sec:Transmission}, we describe our unified transmission strategy. Section~\ref{Sec:Outerbound} is dedicated to the proof of the optimality of our scheme. Section~\ref{Sec:Conclusion} concludes the paper and describes several interesting future directions.

%%%%%%%%%%%%%%%%%%%%%%%%%%%%%%%%%%%%%%%%%%%%%%%%%%%
\section{System Model and Main Results}
\label{Sec:System}

We consider a wireless packet network in which multiple transmitter-receiver pairs wish to communicate with each other. We focus on two nearby transmitter-receiver pairs.  The two pairs are denoted by ${\sf Tx}_1$-${\sf Rx}_1$ and ${\sf Tx}_2$-${\sf Rx}_2$, see Fig.~\ref{Fig:WLAN}. Transmitter $1$ has $m_1$ packets with corresponding physical layer codewords of length $\tau$ denoted by $\vec{a}_1, \vec{a}_2, \ldots, \vec{a}_{m_1}$ and wishes to communicate them to receiver $1$. In this paper, we assume that the mapping from the packets to their corresponding physical layer codewords is fixed, and some point-to-point coding strategy (\emph{e.g.}, LDPC codes, Reed-Solomon codes, etc) is used for the mapping. The details of this mapping does not affect the scheme presented in this paper. The only important issue is the existence of a threshold $\gamma$, such that if codeword $\vec{a}_i$ is transmitted and is received with signal-to-interference-plus-noise ratio of above $\gamma$, the receiver should be able to decode it with high probability. Note that the threshold $\gamma$ depends on how much redundancy is incorporated when encoding the packets (\emph{i.e.} the coding rate). Similarly, transmitter $2$ has $m_2$ packets with corresponding physical layer codewords of length $\tau$ denoted by $\vec{b}_1, \vec{b}_2, \ldots, \vec{b}_{m_2}$ for receiver $2$.

We divide the interference at receiver $i$ into two parts: $(1)$ from the nearby transmitter $\bar{i}$ where $\bar{i} = 3-i$; and $(2)$ from the remaining transmitters in the network. We denote the aggregate interference from the remaining transmitters plus noise at receiver $i$ at time $t$ by $\vec{z}_i(t)$ where $t$ denotes a time slot of length $\tau$ which is enough for transmitting a packet. Therefore, if ${\sf Tx}_1$ and ${\sf Tx}_2$ send packets $\vec{a}$ and $\vec{b}$ at time $t$ respectively, the received signal at receiver $i$ will be given by\footnote{We assume that $g_{ji}(t)$'s are drawn from some continuous distribution (Rayleigh distribution for instance) and are possibly correlated across time.}
\begin{align}
\vec{y}_i(t) = g_{1i}(t) \vec{a} + g_{2i}(t) \vec{b} + \vec{z}_i(t).
\end{align}

We define the signal-to-interference-plus-noise ratio of link $ji$ at ${\sf Rx}_i$, $i,j \in \{ 1, 2\}$, as:
\begin{align}
\mathrm{SINR}_{ji} = 10 \log_{10} \left( \frac{P|g_{ji}|^2}{\mathbb{E}\left[ \vec{z}_i(t) \right] + P |g_{\bar{j}i}|^2} \right),
\end{align}
where $P$ is the average transmit power constraint. Furthermore, define the signal-to-noise ratio (SNR) of link $ji$ at ${\sf Rx}_i$, $i,j \in \{ 1, 2\}$, as:
\begin{align}
\mathrm{SNR}_{ji} = 10 \log_{10} \left( \frac{P|g_{ji}|^2}{\mathbb{E}\left[ \vec{z}_i(t) \right]} \right).
\end{align}

Comparing the definition of $\mathrm{SINR}_{ji}$ and $\mathrm{SNR}_{ji}$, we see that in $\mathrm{SNR}_{ji}$ we do not consider the interference from ${\sf Tx}_{\bar{j}}$.

\begin{figure}[t]
\centering
\subfigure[State~1:~$\mathrm{SINR}_{11} \ge \gamma$]{\includegraphics[height = 1.8 cm]{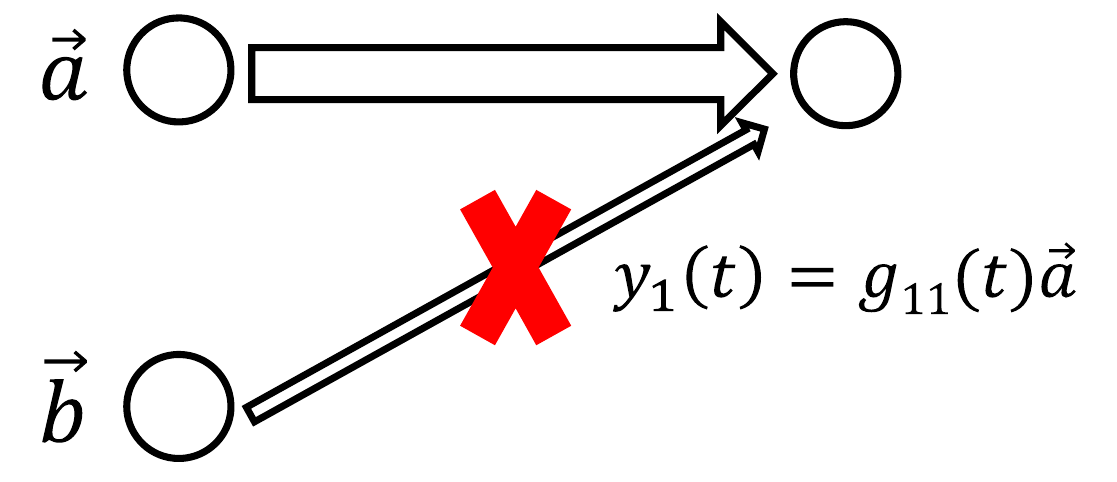}}
\hspace{.12in}
\subfigure[State~2:~$\mathrm{SINR}_{21} \ge \gamma$]{\includegraphics[height = 1.8 cm]{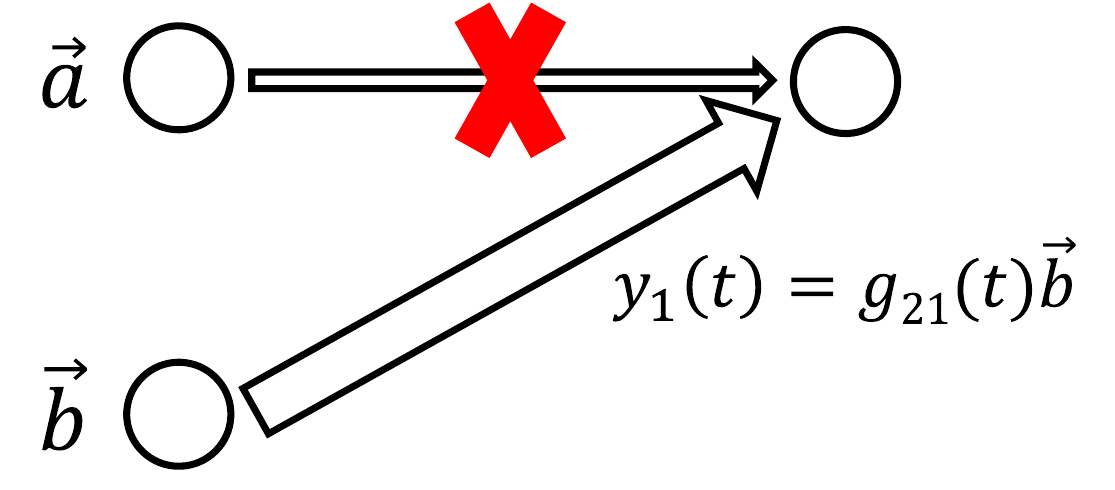}}
% \hspace{.14in}
\subfigure[State~3:~$\mathrm{SINR}_{i1} < \gamma, \mathrm{SNR}_{i1} \ge \gamma$]{\includegraphics[height = 1.8 cm]{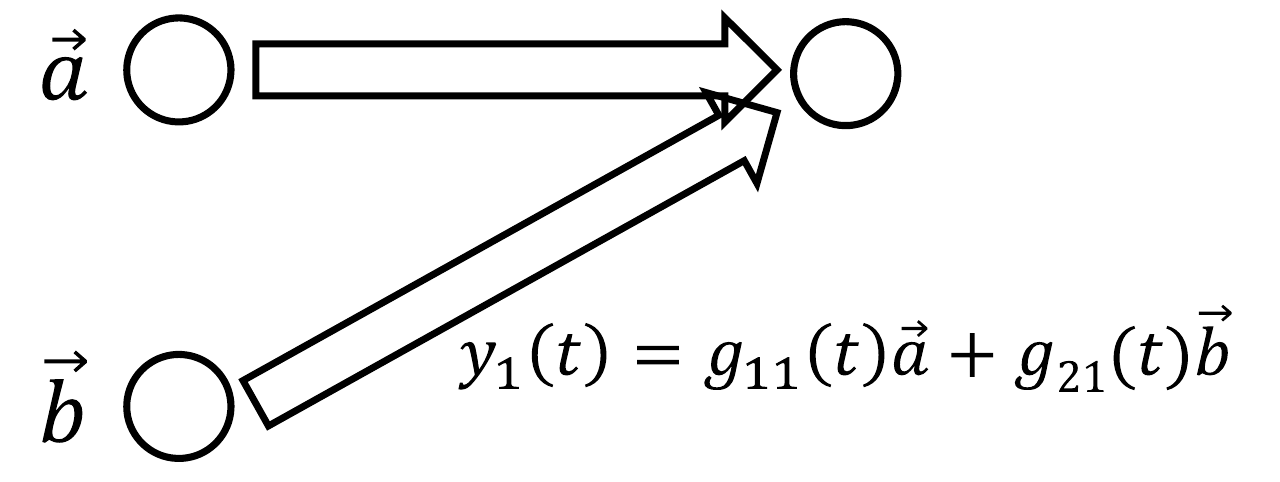}}
\hspace{-.15in}
\subfigure[State~4]{\includegraphics[height = 1.8 cm]{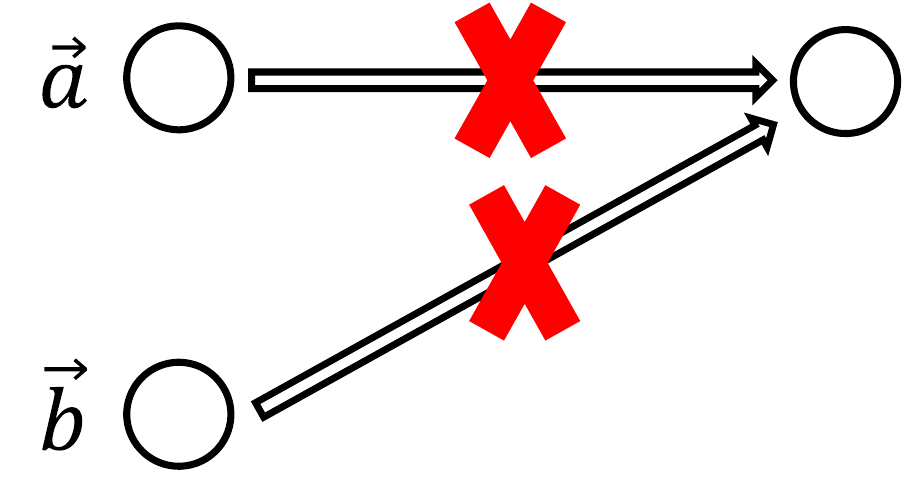}}
\caption{Based on the SINR and SNR values of different links at each time $t$, we have four states.}\label{Fig:OnOff}
\vspace{-5mm}
\end{figure}

Suppose at time $t$, ${\sf Tx}_1$ and ${\sf Tx}_2$ send packets $\vec{a}$ and $\vec{b}$ respectively. Based on the SINR and SNR values of different links at each time $t$, we have one of the following states at any of the receivers, say ${\sf Rx}_1$:
\begin{itemize}

\item State~1~($\mathrm{SINR}_{11} \ge \gamma$): In this state the SINR of the desired packet (\emph{i.e.} $\vec{a}$) at ${\sf Rx}_1$ is above the threshold, and hence, it can be decoded correctly. As shown in Fig.~\ref{Fig:OnOff}(a), this state is as if there is no interference at ${\sf Rx}_1$, and the packet can be decoded properly. 

\item State~2~($\mathrm{SINR}_{21} \ge \gamma$): Similar to State~1, but in this case the SINR of the interfering packet (\emph{i.e.} $\vec{b}$) at ${\sf Rx}_1$ is above the threshold, and hence, it can be decoded correctly, see Fig.~\ref{Fig:OnOff}(b).

\item State~3~($\mathrm{SINR}_{i1} < \gamma$ but $\mathrm{SNR}_{i1} \ge \gamma$ for $i=1,2$): This state corresponds to the scenario that the SINRs of both packets ($\vec{a}$ and $\vec{b}$) are below the threshold at ${\sf Rx}_1$, however, the individual links are strong. Thus, the receiver obtains a linear combination of the packets as depicted in Fig.~\ref{Fig:OnOff}(c). In this case, the receiver cannot decode the packets, however, it stores the signal as the weighted linear combination of the packets. 

% In this case, although the receiver cannot decode the packets, it can store the analog received signal corresponding to the combination of the packets.

\item State~4: In any other scenario, ${\sf Rx}_1$ discards the received signal, see Fig.~\ref{Fig:OnOff}(d).

\end{itemize}

At each time $t$, we define two binary parameters at ${\sf Rx}_1$, $\alpha_{11}(t)$ and $\alpha_{21}(t)$, to determine which state happens:
\begin{align}
\mathrm{State~1} \rightarrow \left( \alpha_{11}(t) = 1, \alpha_{21}(t) = 0 \right), \nonumber \\
\mathrm{State~2} \rightarrow \left( \alpha_{11}(t) = 0, \alpha_{21}(t) = 1 \right), \nonumber \\
\mathrm{State~3} \rightarrow \left( \alpha_{11}(t) = 1, \alpha_{21}(t) = 1 \right), \nonumber \\
\mathrm{State~4} \rightarrow \left( \alpha_{11}(t) = 0, \alpha_{21}(t) = 0 \right).
\end{align}

Based on the definitions, we can capture these cases, through the following abstraction of the physical layer at ${\sf Rx}_1$:
\begin{align}
\label{eq:alphaij}
\vec{y}_1(t) = \alpha_{11}(t) g_{11}(t) \vec{a} + \alpha_{21}(t) g_{21}(t) \vec{b},
\end{align}
for instance, in Case~1 we have $\left( \alpha_{11}(t) = 1, \alpha_{21}(t) = 0 \right)$ and
\begin{align}
\label{eq:case1}
\vec{y}_1(t) = g_{11}(t) \vec{a},
\end{align}
and in Case~3 we have $\left( \alpha_{11}(t) = 1, \alpha_{21}(t) = 1 \right)$ and
\begin{align}
\label{eq:case3}
\vec{y}_1(t) = g_{11}(t) \vec{a} + g_{21}(t) \vec{b}.
\end{align}
Similarly, we define $\alpha_{12}(t)$ and $\alpha_{22}(t)$, and we will have four similar states for ${\sf Rx}_2$. Therefore, from the point of view of a receiver, we have four states, and a total of $16$ distinct cases for the channel configurations as in Table~\ref{table:16cases}.

We represent the channel-state at time instant $t$ by the quadruple 
\begin{align}
\alpha(t) = (\alpha_{11}(t), \alpha_{12}(t), \alpha_{21}(t), \alpha_{22}(t)).
\end{align}

The channel-state varies over time due to: $(1)$ mobility of the two transmitter-receiver pairs and fading, and $(2)$ time-varying aggregate interference from other users. We assume that each transmitter is aware of the channel-state information with some delay which for simplicity assumed to be one transmission block (\emph{i.e.} $\tau$). Hence, at time instant $t$, each transmitter knows $\alpha^{t-1} = \left( \alpha(\ell) \right)_{\ell = 1}^{t-1}$. We refer to this model of knowledge as the Delayed Channel-State Information at the Transmitters (Delayed CSIT) model. We further assume that the receivers have a delayed knowledge of the channel-state and the channel gains.

In general, $\alpha_{ji}(t)$'s are correlated across time and with respect to each other. This correlation allows us to predict future and improve the throughput that way. However, in this paper, we aim to focus on the gains from the stored analog signals at the receivers. Therefore, in order to isolate ourselves from the benefit of predicting the future,  we assume that $\alpha_{ji}(t)$'s vary as i.i.d. Bernoulli random variables $\mathcal{B}(p)$ for $0 \leq p \leq 1$. For convenience, we denote a linear combination of packets $\vec{a}$ and $\vec{b}$ by $L(\vec{a},\vec{b})$.

Based on this model, we define the achievable throughput region of ${\sf Tx}_1$ and ${\sf Tx}_2$ as follows. Consider the scenario in which ${\sf Tx}_i$ wishes to reliably communicate $m_i$ packets to ${\sf Rx}_i$ during $n$ uses of the channel, $i = 1,2$. We assume that the packets and the channel gains are {\it mutually} independent. Receiver ${\sf Rx}_i$ is only interested in packets from ${\sf Tx}_i$, and it will recover (decode) them using the received signals $\vec{y}_i^n$ and the knowledge of the channel state information.

At each time instant, transmitter $i$ creates a linear combination of the $m_i$ packets it has for receiver $i$ by choosing a precoding vector $\vec{v}_i(t) \in \mathbb{R}^{1 \times m_i}$, $i=1,2$. Then, the transmit signals at time $t$ at ${\sf Tx}_1$ and ${\sf Tx}_2$ are given by $\vec{v}_1(t) \mathbf{A}$ and $\vec{v}_2(t) \mathbf{B}$
%\begin{align}
%\vec{v}_1(t) \mathbf{A}, \qquad \vec{v}_2(t) \mathbf{B},
%\end{align}
respectively, where $\mathbf{A} = \left[ \vec{a}_1, \vec{a}_2, \ldots, \vec{a}_{m_1} \right]^{\top}$, and $\mathbf{B} = \left[ \vec{b}_1, \vec{b}_2, \ldots, \vec{b}_{m_2} \right]^{\top}$. We impose the following constraints on $\vec{v}_1(t)$ and $\vec{v}_2(t)$ to satisfy the power constraint at the transmitters:
\begin{align}
||\vec{v}_1(t)||, ||\vec{v}_2(t)|| \leq 1,
\end{align}
where $||.||$ represents the Euclidean norm. Due to the Delayed-CSIT assumption, $\vec{v}_i(t)$ is only a function of $\alpha^{t-1}$. The received signal of receiver $i$ at time $t$, can be represented by
\begin{align}
\vec{y}_i(t) = \alpha_{1i}(t) g_{1i}(t) \vec{v}_1(t) \mathbf{A} + \alpha_{2i}(t) g_{2i}(t) \vec{v}_2(t) \mathbf{B}.
\end{align}
We denote the overall precoding matrix of transmitter $i$ by $\mathbf{V}_i^n \in \mathbb{R}^{n \times m_i}$, where the $t^{\mathrm{th}}$ row of $\mathbf{V}_i^n$ is $\vec{v}_i(t)$. Furthermore, let $\mathbf{G}_{ij}^n$ be an $n \times n$ diagonal matrix where the $t^{\mathrm{th}}$ diagonal element is $\alpha_{ij}(t) g_{ij}(t)$, $i,j = 1,2$. Thus, we can right the output at receiver $i$ as
\begin{align}
\label{eq:PHYLayer}
\vec{y}_i^n = \mathbf{G}_{1i}^n \mathbf{V}_1^n \mathbf{A} + \mathbf{G}_{2i}^n \mathbf{V}_2^n \mathbf{B}, \quad i = 1,2.
\end{align}

We denote the interference subspace at receiver $i$ by $\mathcal{I}_i$ and is given by
\begin{align}
\mathcal{I}_i = \mathrm{colspan}\left( \mathbf{G}_{\bar{i}i} \mathbf{V}_{\bar{i}} \right), \quad i = 1,2,
\end{align}
where $\mathrm{colspan}$ of a matrix represents the sub-space spanned by its column vectors, and let $\mathcal{I}_i^c$ denote the subspace orthogonal to $\mathcal{I}_i$. Then in order for decoding to be successful at receiver $i$, it should be able to create $m_i$ linearly independent equations that are solely in terms of its intended packets. Mathematically speaking, this means that the image of $\mathrm{colspan}\left( \mathbf{G}_{ii}^n \mathbf{V}_{i}^n \right)$ on $\mathcal{I}_i^c$ should have the same dimension as  $\mathrm{colspan}\left( \mathbf{V}_{i}^n \right)$ itself. More precisely, we require
\begin{align}
\label{eq:decodability}
\mathrm{dim} & \left( \mathrm{Proj}_{\mathcal{I}_i^c}~\mathrm{colspan}\left( \mathbf{G}_{ii}^n \mathbf{V}_{i}^n \right) \right) \nonumber \\
& = \mathrm{dim}\left( \mathrm{colspan}\left( \mathbf{V}_{i}^n \right) \right) = m_i,  \quad i = 1,2.
\end{align}

\begin{figure}[b]
\centering
\includegraphics[height = 5 cm]{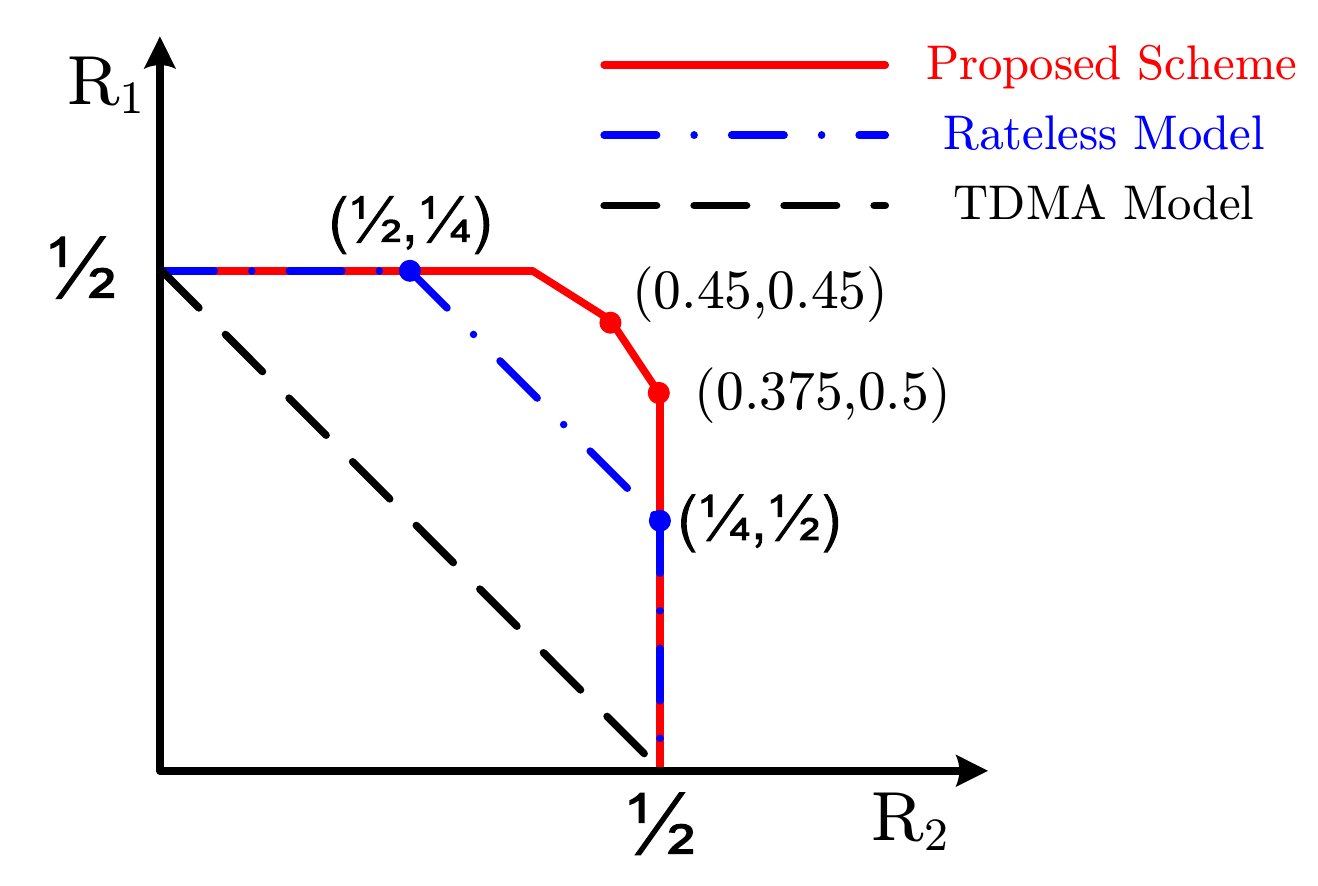}
\caption{A comparison of our proposed scheme with TDMA model and rateless model for $p = 1/2$.\label{fig:Comparison}}
\end{figure}

We say that a throughput tuple of $\left( R_1,R_2 \right) = \left( m_1/n, m_2/n \right)$ is achievable, if there exists a choice of $\mathbf{V}_1^n$ and $\mathbf{V}_2^n$, such that (\ref{eq:decodability}) is satisfied for $i=1,2$ with probability $1$. The throughput region, $\mathcal{C}$, is then the closure of all achievable throughput tuples $\left( R_1,R_2 \right)$.

As benchmarks for comparison, we consider the following schemes.
\begin{enumerate}

\item Transmission under time division model (TDMA): In this scheme, at any given time only one transmitter can communicate. When a transmitter talks during its allocated interval, it has the opportunity to communicate with its receiver only a fraction $p$ of the times. That is due to the fact that only a faction $p$ of the time slots, 
the corresponding direct link is on, meaning the SINR is above the required threshold ($\alpha_{ij}(t) \sim \mathcal{B}(p)$). Therefore, the achievable throughput in this model is $p$. In Fig.~\ref{fig:Comparison}, we have plotted the achievable throughput region of the TDMA model for $p=1/2$.

\item Transmission under rateless codes: In this scheme, transmitters try to provide enough equations (by using rateless codes for instance) to the receivers such that each receiver can recover all packets (not just the ones intended for it). This is one way to take advantage of the available analog signals at the receivers. Since each receiver obtains useful information only $(1-p)^2$ of the times, therefore, the sum throughput would be $(1-p)^2$. In Fig.~\ref{fig:Comparison}, we have plotted the achievable throughput region of the rateless model for $p=1/2$. As we can see, the sum throughput is improved from $1/2$ to $3/4$ which indicates an improvement of $50\%$.

\end{enumerate}

In this paper, we study the optimal throughput region for the physical layer model described in (\ref{eq:PHYLayer}). In particular, we develop a new transmission strategy that incorporates two novel coding opportunities at the transmitters for interference management, and goes well beyond the aforementioned benchmark schemes. We also prove the optimality of our scheme in the context of a network with two transmitter-receiver pairs that can coordinate for interference management as discussed above, hence, characterizing the throughout region of this network as follows.

\begin{theorem}
\label{THM:Main}
The throughput region, $\mathcal{C}$, of the network in which two transmitter-receiver pairs can coordinate as discussed above, is as follows:
\begin{equation}
\label{eq:Main}
\mathcal{C} =
\left\{ \begin{array}{ll}
\vspace{1mm} 0 \leq R_i \leq p, & i = 1,2, \\
R_i + (2-p) R_{\bar{i}} \leq p(2-p)^2, & i = 1,2,
\end{array} \right.
\end{equation}
where we have assumed $\alpha_{ji}(t)$'s in (\ref{eq:alphaij}) are distributed as i.i.d. $\mathcal{B}(p)$ random variables for $0 \leq p \leq 1$.
\end{theorem}

We have plotted the achievable throughput region of our proposed scheme for $p = 1/2$ in Fig.~\ref{fig:Comparison}. As we can see, the sum throughput is $0.9$ which is well beyond the two other schemes. In fact, it indicates an improvement of $90\%$ compared to the TDMA model, and an improvement of $20\%$ compared to the rateless model.

In Section~\ref{Sec:Transmission}, we describe our unified transmission strategy for this theorem, and in Section~\ref{Sec:Outerbound}, by deriving an outer-bound, we prove the optimality of our scheme. 
%%%%%%%%%%%%%%%%%%%%%%%%%%%%%%%%%%%%%%%%%%%%%%%%%%%

\section{Coding Opportunities}
\label{Sec:Opportunities}

In this section, we show how the stored analog received signals from the past can be utilized to better manage interference. In particular, we illustrate two novel coding opportunities that later in Section~\ref{Sec:Transmission}, we use to design the optimal transmission strategy for the network with two transmitter-receiver pairs (as stated in Theorem~\ref{THM:Main}). We should mention that although we mainly talk about the network with two transmitter-receiver pairs, the ideas and concepts we talk about here are general and not limited to this specific network.

We start by discussing a conventional approach of utilizing the previously received signals at the receivers to better manage interference by the transmitters, namely {\it packet repetition coding}. This technique can be best explained through the following example.
\begin{figure}[h]
\vspace{-5mm}
\centering
\subfigure[]{\includegraphics[height = 2.75 cm]{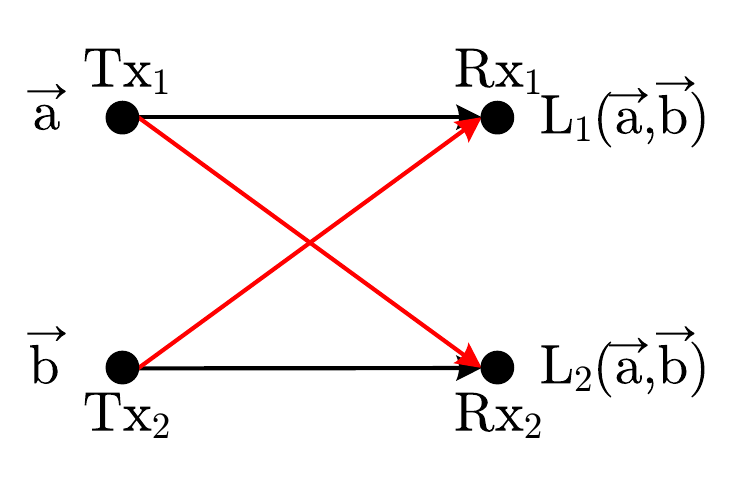}}
\hspace{0.15 in}
\subfigure[]{\includegraphics[height = 2.75 cm]{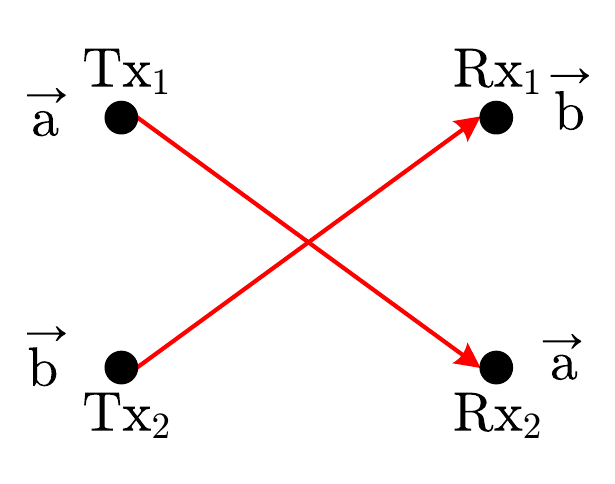}}
\caption{$(a)$ We only need to retransmit one of the packets, say $\vec{a}$, to both receivers, and we refer to this packet as a packet of {\it common interest}; $(b)$ packets $\vec{a}$ and $\vec{b}$ are already decoded at the unintended receivers, we can update their status to {\it interference-free packets}.\label{Fig:Conventional}}
\vspace{-2.5mm}
\end{figure}

Assume that at a time instant, ${\sf Tx}_1$ and ${\sf Tx}_2$ send packets $\vec{a}$ and $\vec{b}$ respectively, and each receiver is in State~3 as described in Section~\ref{Sec:System}, see Fig.~\ref{Fig:Conventional}(a). Hence, each receiver obtains a linear combination of the packets. Now, instead of discarding the received signals, and retransmission of both $\vec{a}$ and $\vec{b}$, we can store the analog equations at the receivers. We then only retransmit one of them, say $\vec{a}$, to both receivers. This way, ${\sf Rx}_2$ gets $\vec{a}$, and it can use $\vec{a}$ and $L_2\left(\vec{a},\vec{b}\right)$ to cancel interference and decode the desired packet (\emph{i.e.} $\vec{b}$). So in a sense, after occurrence of the configuration in Fig.~\ref{Fig:Conventional}(a), we can update the status of packets $\vec{a}$ and $\vec{b}$ as follows. Packet $\vec{a}$ becomes a packet of {\it common interest} to both receivers. Packet $\vec{a}$ joins queue $Q_{1 \rightarrow \{ 1, 2\} }$ that represents the packets at ${\sf Tx}_1$ that are of interest of both receivers. Packet $\vec{b}$ becomes virtually delivered (since when $\vec{a}$ is delivered to ${\sf Rx}_2$, packet $\vec{b}$ can be decoded using SIC).

\begin{figure}[h]
\vspace{-2.5mm}
\centering
\includegraphics[height = 2.75 cm]{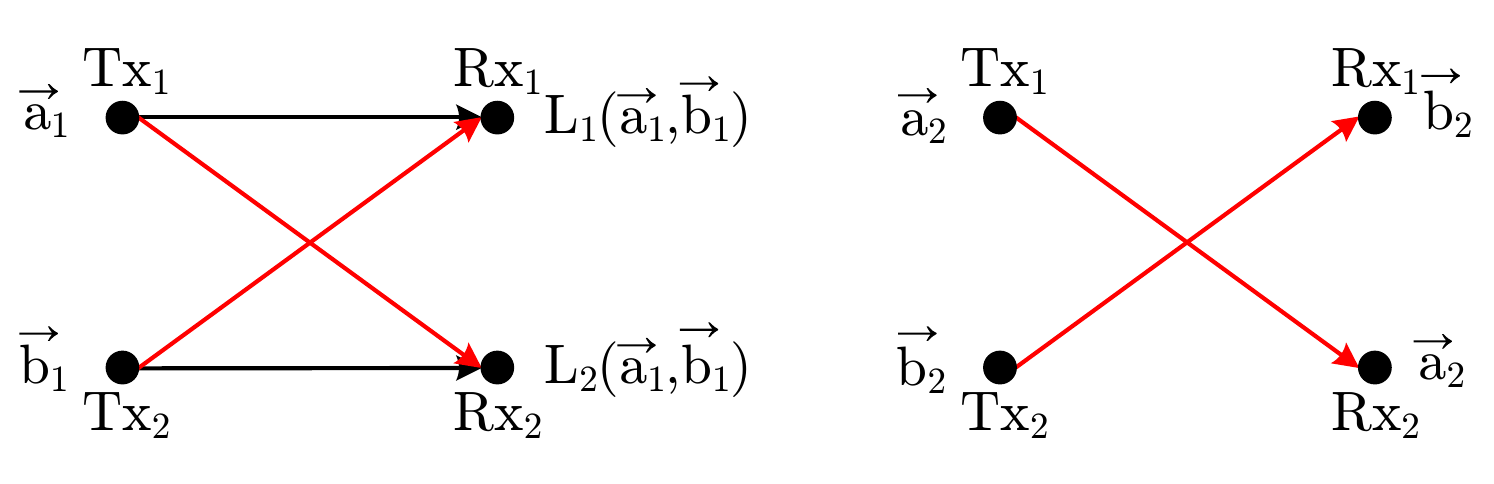}
\caption{Coding opportunity Type-I ({\it packet delivery with side information}): It is sufficient to provide $\left( \vec{a}_1 + \vec{a}_2 \right)$ and $\left( \vec{b}_1 + \vec{b}_2 \right)$ to both receivers.\label{Fig:CodingTypeI}}
\vspace{-2.5mm}
\end{figure}

\begin{figure*}[t]
\centering
\includegraphics[height = 2.75 cm]{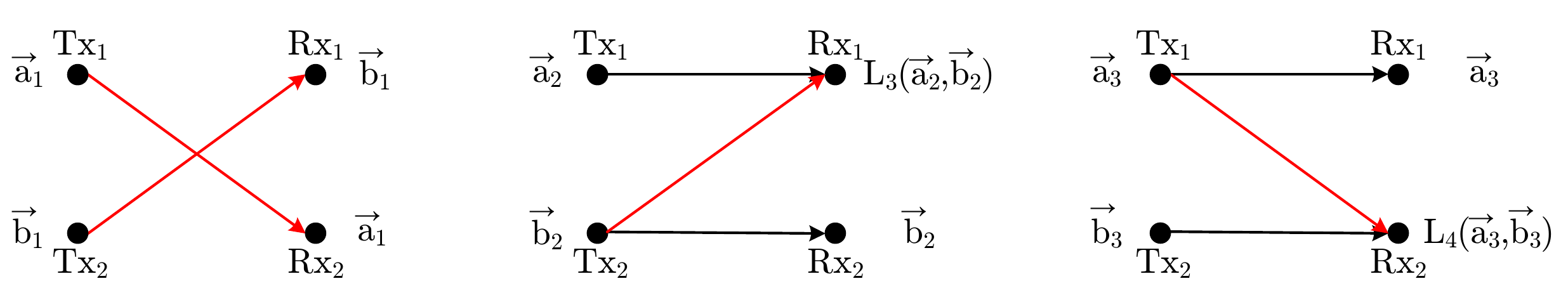}
\caption{Coding opportunity Type-II ({\it interference delivery with side information}): It is sufficient to provide $\left( \vec{a}_1 + \vec{a}_3 \right)$ and $\left( \vec{b}_1 + \vec{b}_2 \right)$ to both receivers.\label{Fig:CodingTypeII}}
\vspace{-2.5mm}
\end{figure*}

Similarly, suppose at a time instant, ${\sf Tx}_1$ and ${\sf Tx}_2$ send packets $\vec{a}$ and $\vec{b}$ respectively, and each receiver is in State~2 as described in Section~\ref{Sec:System}, see Fig.~\ref{Fig:Conventional}(b). In this scenario, each receiver can successfully decode the packet of the unintended receiver, but not its own. However, since the unintended receivers have packets $\vec{a}$ and $\vec{b}$, from now on these packets cannot create interference. Therefore, we can update their status to {\it interference-free packets}, and move them to queues $Q_{1 \rightarrow 1|2}$ and $Q_{2 \rightarrow 2|1}$ respectively (queue $Q_{i \rightarrow i|\bar{i}}$ represents the packets at transmitter $i$ that are required by receiver $i$ but available at receiver $\bar{i}$, $i=1,2$).

Interference-free packets seem very appealing. At first glance, one might think that simply retransmitting these packets would be the best approach. However, perhaps surprisingly, we show that by combining them with other packets, we can achieve even higher throughput. In what follows, we show that by innovative coding of interference-free packets and packets of common interest, one can do much better.  In particular, we identify two coding opportunities as follows.

\noindent $\bullet$ {\bf Coding opportunity Type-I} ({\it packet delivery with side information}): Consider two time instants in which each one of the transmitters send two packets (denoted by $\vec{a}_1,\vec{a}_2$ at ${\sf Tx}_1$ and $\vec{b}_1,\vec{b}_2$ at ${\sf Tx}_2$) and the channel configurations are as shown in Fig.~\ref{Fig:CodingTypeI}. As mentioned before, either one of $\vec{a}_1$ or $\vec{b}_1$ can be considered as a ``packet of common interest'', and $\vec{a}_2$ and $\vec{b}_2$ are ``interference-free packets''. Under the conventional approach described above, we have to retransmit $\vec{a}_1$ (or $\vec{b}_1$), $\vec{a}_2$ and $\vec{b}_2$. However, there is a more efficient way of delivering the packets by the following coding idea. We observe that providing $\left( \vec{a}_1 + \vec{a}_2 \right)$ and $\left( \vec{b}_1 + \vec{b}_2 \right)$ to both receivers is sufficient to recover the packets
%\footnote{To be precise, in order to satisfy the power constraint at the transmitter, we need to normalize this summation as $\left( \vec{a}_1 + \vec{a}_2 \right)/\sqrt{2}$, however, for convenience we do not carry $1/\sqrt{2}$ along.}. 
For instance, if $\left( \vec{a}_1 + \vec{a}_2 \right)$ and $\left( \vec{b}_1 + \vec{b}_2 \right)$ are available at ${\sf Rx}_1$, it can subtract $\vec{b}_2$ to recover $\vec{b}_1$, then using $\vec{b}_1$ and $L_1\left( \vec{a}_1, \vec{b}_1 \right)$ it can obtain $\vec{a}_1$; finally, using $\vec{a}_1$ and $\left( \vec{a}_1 + \vec{a}_2 \right)$ it can recover $\vec{a}_2$. Similar argument holds for the other receiver. Indeed, the linear combination $\left( \vec{a}_1 + \vec{a}_2 \right)$ available at ${\sf Tx}_1$, and $\left( \vec{b}_1 + \vec{b}_2 \right)$ available at ${\sf Tx}_2$, are packets of common interest and they join $Q_{1 \rightarrow \{ 1, 2\}}$ and $Q_{2 \rightarrow \{ 1, 2\}}$ respectively.

To clarify why coding opportunity Type-I is useful, we consider a numerical example. In the scenario where $\alpha_{ij}(t)$'s are $\mathcal{B}(1/2)$ random variables, in average, two interference-free packets can be communicated in two time instants or one packet per time instant. On the other hand as we will see later, in average, one packet of common interest can be communicated in $4/3$ time instants. Thus, in average, we can deliver $\vec{a}_1,\vec{b}_1,\vec{a}_2$ and $\vec{b}_2$ in $10/3$ time instants. However, after coding opportunity Type-I and delivering two packets of common interest, in average, we are in fact recovering $4$ packets in $8/3$ time instants which indicates $20\%$ improvement.

\noindent $\bullet$ {\bf Coding opportunity Type-II} ({\it interference delivery with side information}): Consider three time instants in which each one of the transmitters send three packets (denoted by $\vec{a}_1,\vec{a}_2,\vec{a}_3$ at ${\sf Tx}_1$ and $\vec{b}_1,\vec{b}_2,\vec{b}_3$ at ${\sf Tx}_2$) and then through Delayed-CSIT, the transmitters realize that the channel configurations were as shown in Fig.~\ref{Fig:CodingTypeII}. Now, under the conventional approach described above, we can move $\vec{a}_1$ and $\vec{b}_1$ to $Q_{1 \rightarrow 1|2}$ and $Q_{2 \rightarrow 2|1}$ respectively; and we can retransmit $\vec{a}_2$ and $\vec{b}_3$. However, the following coding idea provides a more efficient way of delivering the packets. The main idea is to take advantage of packets $\vec{a}_2$ and $\vec{b}_3$ which are already available at their receivers. Transmitters $1$ and $2$ can respectively create two coded packets $\left( \vec{a}_1 + \vec{a}_3 \right)$ and $\left( \vec{b}_1 + \vec{b}_2 \right)$. Now note that if $\left( \vec{a}_1 + \vec{a}_3 \right)$ and $\left( \vec{b}_1 + \vec{b}_2 \right)$ are available at ${\sf Rx}_1$, it can subtract $\vec{b}_1$ to recover $\vec{b}_2$, then using $\vec{b}_2$ and $L_1\left( \vec{a}_2, \vec{b}_2 \right)$ it can obtain $\vec{a}_2$; finally, using $\vec{a}_3$ and $\left( \vec{a}_1 + \vec{a}_3 \right)$ it can recover $\vec{a}_1$. Similar argument holds for the other receiver. Indeed, the linear combination $\left( \vec{a}_1 + \vec{a}_3 \right)$ available at ${\sf Tx}_1$, and $\left( \vec{b}_1 + \vec{b}_2 \right)$ available at ${\sf Tx}_2$, are ``packets of common interest'' and they join $Q_{1 \rightarrow \{ 1, 2\}}$ and $Q_{2 \rightarrow \{ 1, 2\}}$ respectively.

Now, a natural question arises: ``Are there more efficient coding strategies?''. We will next answer this question for the network with two transmitter-receiver pairs. In particular, in Section~\ref{Sec:Transmission}, we propose an efficient way of combining and concatenating the coding opportunities presented in this section, and in Section~\ref{Sec:Outerbound}, we prove the optimality of our scheme. Hence, in the context of networks with two transmitter-receiver pairs, the coding opportunities discussed in this section are indeed sufficient to achieve the optimal throughput.

%%%%%%%%%%%%%%%%%%%%%%%%%%%%%%%%%%%%%%%%%%%%%%%%%%%
\section{Optimal Concatenation of Coding Opportunities}
\label{Sec:Transmission}

In this section, we describe how the key ideas developed in Section~\ref{Sec:Opportunities}, can be systematically utilized and concatenated to achieve the optimal throughput in a network in which two transmitter-receiver pairs can coordinate for interference management. In fact, we show how to achieve the throughput region given in Theorem~\ref{THM:Main} for $p = 1/2$. The transmission strategy for general value of $p$ is similar and is omitted here. The main ideas required to extend this result for general value of $p$ can be found in~\cite{vahid2013capacity} Section~V in the context of a two-user binary fading interference channel.

We show that we can achieve a throughput arbitrary close to corner point $\left( 0.45, 0.45 \right)$, see Fig.~\ref{fig:Comparison}. In fact, we show that it is possible to communicate the initial $2m$ packets in 
\begin{align}
n = \left( 20/9 \right) m + O\left( m^{2/3}\right)
\end{align}
time instants\footnote{Throughout the paper whenever we state the number of packets or time instants, say $n$, if the expression is not an integer, then we use the ceiling of that number $\lceil n \rceil$, where $\lceil . \rceil$ is the smallest integer greater than or equal to $n$. Note that since we will take the limit as $m \rightarrow \infty$, this does not change the end results.} with vanishing error probability (as $m \rightarrow \infty$). Therefore achieving corner point $\left( 0.45, 0.45 \right)$ as $m \rightarrow \infty$. The transmission strategy for corner points $\left( 0.375, 0.5 \right)$ and $\left( 0.5, 0.375 \right)$ follows similar principles and is omitted due to space limitations. Our transmission strategy consists of two phases as described below. 

%%%%%%%%%%%%%%%%%%%%%%%%%%%%%%%%%%%%%%%%%%%%%%%%%%%%%%%%%%%%%%%%%%%%%%%%%%%%
%%%%%%%%%%%%%%%%%%%%%%%%%%%%%%%%%%%%%%%%%%%%%%%%%%%%%%%%%%%%%%%%%%%%%%%%%%%%
%%%%%%%%%%%%%%%%%%%%%%%%%%%%%%%%%%%%%%%%%%%%%%%%%%%%%%%%%%%%%%%%%%%%%%%%%%%%

\begin{table}[t]
\caption{All connectivity configurations and status transitions. $\vec{a}$ represents a packet in $Q_{1 \rightarrow 1}$ and $\vec{b}$ represents a packet in $Q_{2 \rightarrow 2}$.}
\centering
\begin{tabular}{| c | c |}

\hline

\includegraphics[height = 1.5cm]{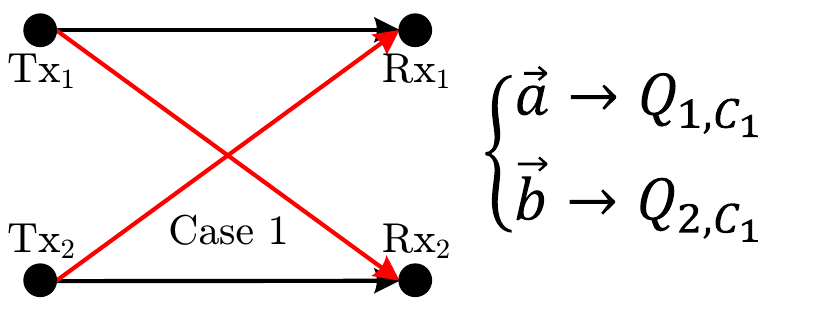}     &  \includegraphics[height = 1.5cm]{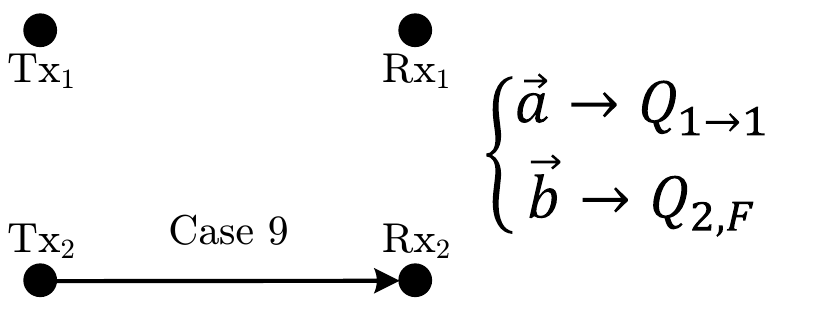}  \\

\hline

\includegraphics[height = 1.4cm]{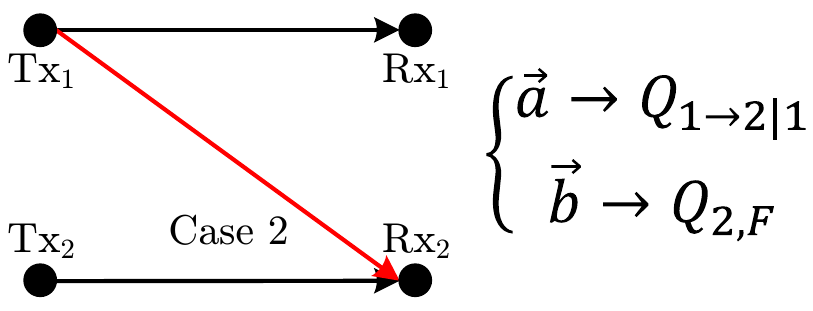}     &  \includegraphics[height = 1.4cm]{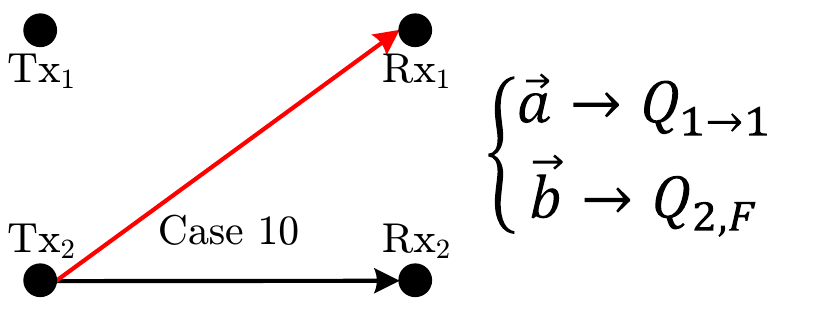}  \\

\hline

\includegraphics[height = 1.4cm]{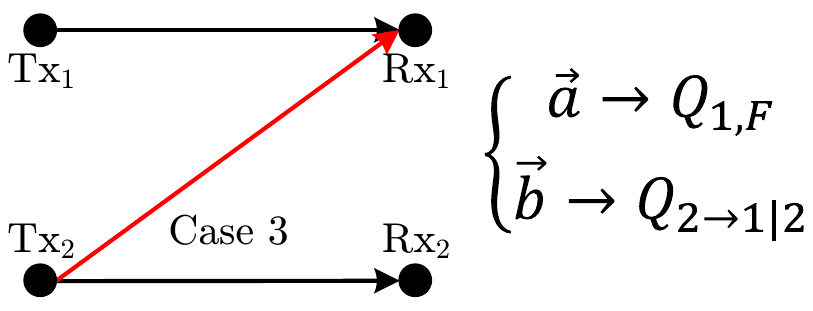}     &  \includegraphics[height = 1.4cm]{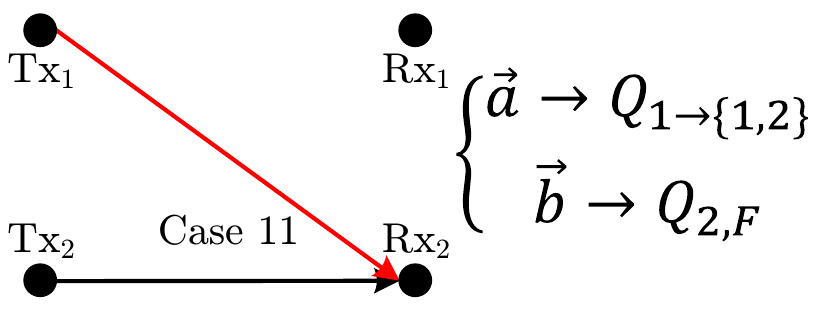}  \\

\hline

\includegraphics[height = 1.4cm]{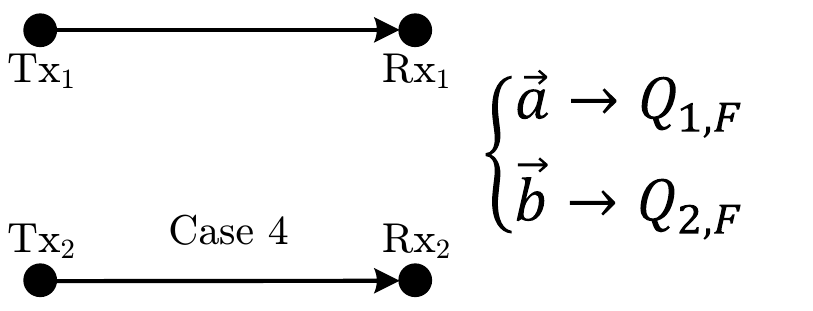}     &  \includegraphics[height = 1.4cm]{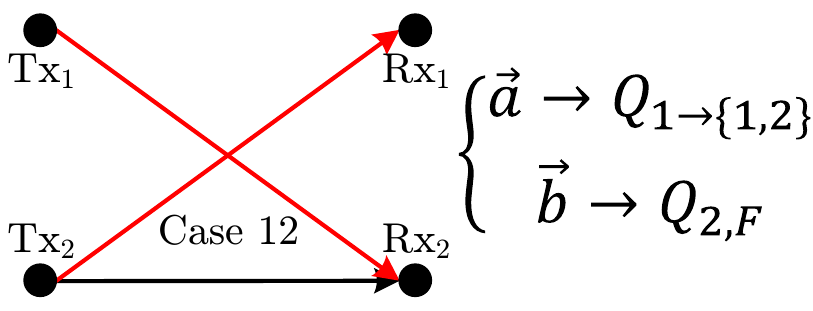}  \\

\hline

\includegraphics[height = 1.4cm]{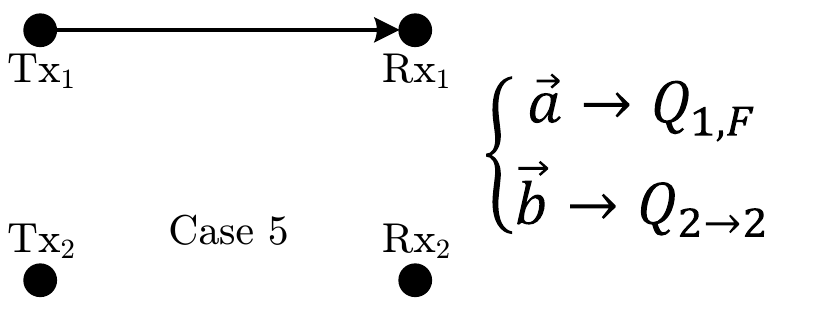}     &  \includegraphics[height = 1.4cm]{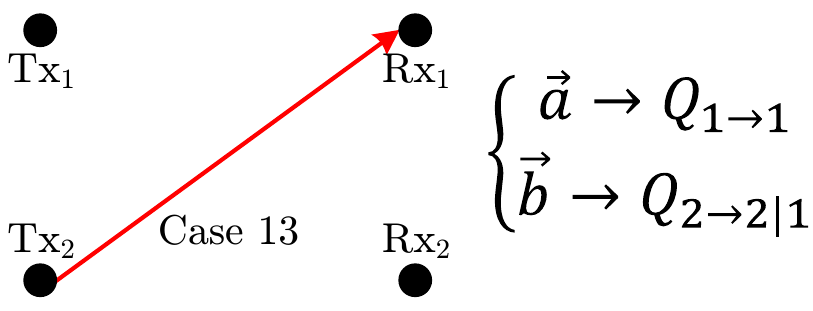}  \\

\hline

\includegraphics[height = 1.4cm]{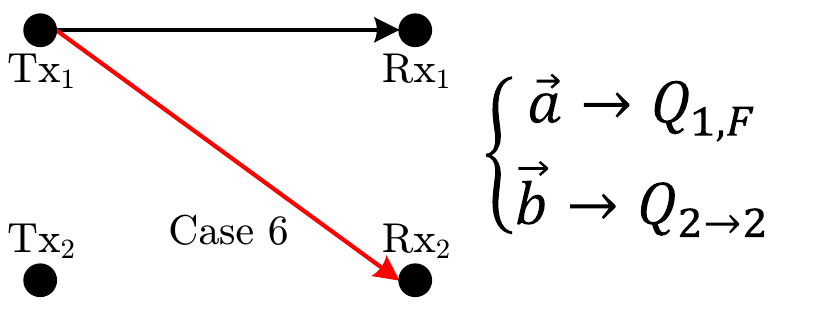}     &  \includegraphics[height = 1.4cm]{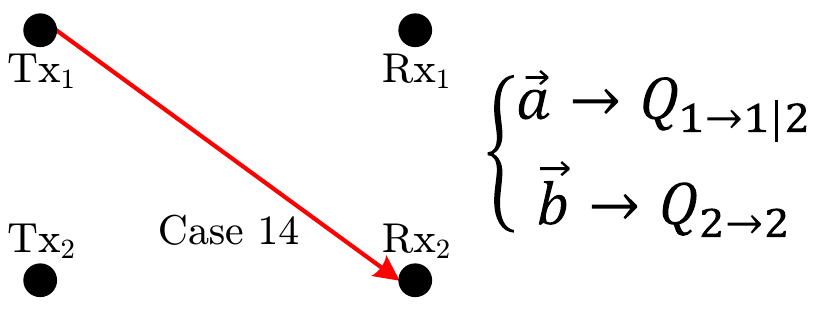}  \\

\hline

\includegraphics[height = 1.4cm]{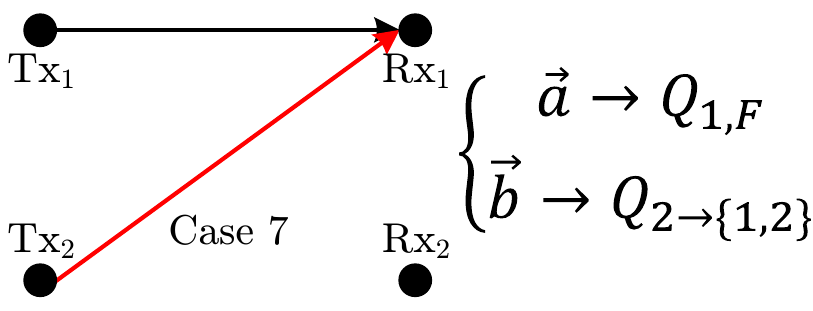}     &  \includegraphics[height = 1.4cm]{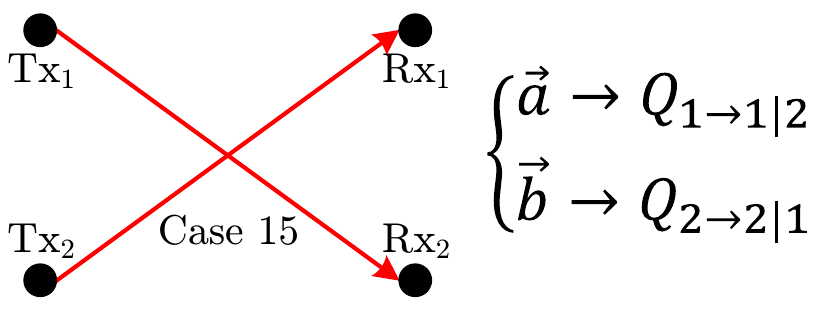}  \\

\hline

\includegraphics[height = 1.4cm]{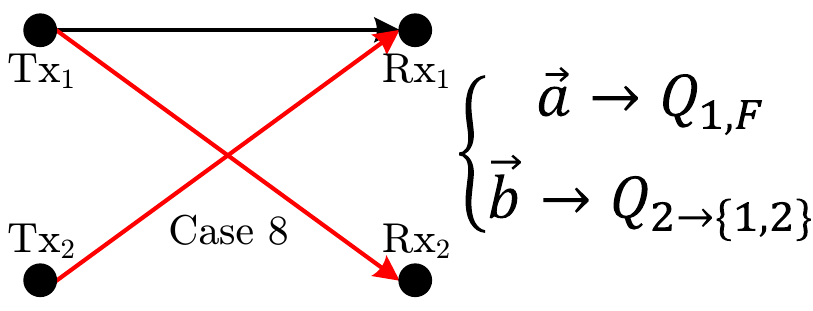}     &  \includegraphics[height = 1.4cm]{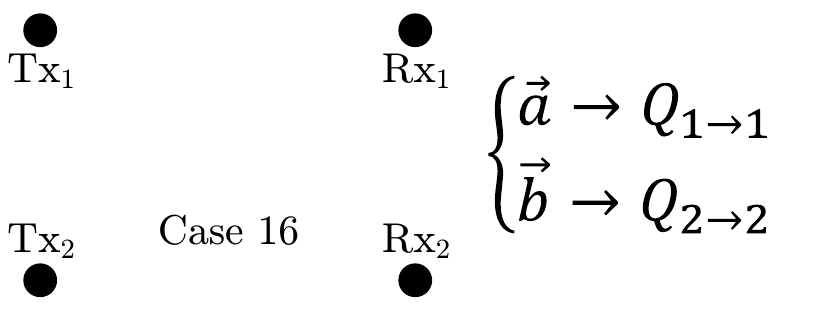}  \\

\hline

\end{tabular}
\label{table:16cases}
\end{table}

%%%%%%%%%%%%%%%%%%%%%%%%%%%%%%%%%%%%%%%%%%%%%%%%%%%%%%%%%%%%%%%%%%%%%%%%%%%%
%%%%%%%%%%%%%%%%%%%%%%%%%%%%%%%%%%%%%%%%%%%%%%%%%%%%%%%%%%%%%%%%%%%%%%%%%%%%
%%%%%%%%%%%%%%%%%%%%%%%%%%%%%%%%%%%%%%%%%%%%%%%%%%%%%%%%%%%%%%%%%%%%%%%%%%%%

\noindent {\bf Phase 1} [uncategorized transmission]: At the beginning of the communication block, we assume that the packets at ${\sf Tx}_i$ are in queue $Q_{i \rightarrow i}$ (the initial state of the packets), $i=1,2$. At each time instant $t$, ${\sf Tx}_i$ sends out a packet from $Q_{i \rightarrow i}$, and this packet will either stay in the initial queue or transition to one of the following possible queues will take place according to the description in Table~\ref{table:16cases}. If at time instant $t$, $Q_{i \rightarrow i}$ is empty, then ${\sf Tx}_i$, $i=1,2$, remains silent until the end of Phase~1.
\begin{enumerate}
\item[(A)] $Q_{i , {\sf C}_1 }$: The packets that at the time of communication, all channel gains were on.
\item[(B)] $Q_{i \rightarrow \{ 1,2 \} }$: The packets that are of common interest of both receivers and do not fall in category (A).  
\item[(C)] $Q_{i \rightarrow i|\bar{i}}$: The packets that are required by ${\sf Rx}_i$ but are available at the unintended receiver ${\sf Rx}_{\bar{i}}$ where $\bar{i} = 3 - i$. A packet is in  $Q_{i \rightarrow i|\bar{i}}$ if ${\sf Rx}_{\bar{i}}$ gets it without interference and ${\sf Rx}_i$ does not get it with or without interference.
\item[(D)] $Q_{i \rightarrow \bar{i}|i}$: The packets that are required by ${\sf Rx}_{\bar{i}}$ but are available at the intended receiver ${\sf Rx}_i$. More precisely, a packet is in  $Q_{i \rightarrow \bar{i}|i}$ if ${\sf Rx}_{i}$ gets the packet without interference and ${\sf Rx}_{\bar{i}}$ gets it with interference.
\item[(E)] $Q_{i \rightarrow F}$: The packets that we consider delivered and no retransmission is required.
\end{enumerate}

More precisely, based on the channel realizations, a total of $16$ possible configurations may occur at any time instant as summarized in Table~\ref{table:16cases}. The transition for each one of the channel realizations is as follows.

\begin{itemize}

\item Case~1~$\left( \caseone \right)$: If at time instant $t$, Case 1 occurs, then each receiver gets a linear combination of the packets that were transmitted. Then, if either of such packets is provided to both receivers then the receivers can recover both. The transmitted packet of ${\sf Tx}_i$ leaves $Q_{i \rightarrow i}$ and joins $Q_{i,{\sf C}_1}$\footnote{In this paper, we assume that the queues are ordered. Meaning that the first packet that joins the queue is placed at the head of the queue and any new packet occupies the next empty position. For instance, suppose there are $\ell$ packets in $Q_{1,C_1}$ and $\ell$ packets in $Q_{2,C_1}$, then the next time Case~1 occurs, the transmitted packet of ${\sf Tx}_i$ is placed at position $\ell + 1$ in $Q_{i,C_1}$, $i=1,2$.}, $i=1,2$. Although we can consider such packets as packets of common interest, we keep them in an intermediate queue for now and as we describe later, we combine them with other packets to create packets of common interest. 

\item Case~2~$\left( \casetwo \right)$: In this case, ${\sf Rx}_1$ has already received its corresponding packet while ${\sf Rx}_2$ has a linear combination of the transmitted packets, see Table~\ref{table:16cases}. As a result, if the transmitted packet of ${\sf Tx}_1$ is provided to ${\sf Rx}_2$, it will be able to decode both. In other words, the transmitted packet from ${\sf Tx}_1$ is available at ${\sf Rx}_1$ and is required by ${\sf Rx}_2$. Therefore, transmitted packet of ${\sf Tx}_1$ leaves $Q_{1 \rightarrow 1}$ and joins $Q_{1 \rightarrow 2|1}$. Note that the packet of ${\sf Tx}_2$ will not be retransmitted since upon delivery of the packet of ${\sf Tx}_1$, ${\sf Rx}_2$ can decode its corresponding packet. Since no retransmission is required, the packet of ${\sf Tx}_2$ leaves $Q_{2 \rightarrow 2}$ and joins $Q_{2,F}$ (the final state of the packets).

\vspace{1mm}

\item Case~3~$\left( \casethree \right)$: This is similar to Case~2 with swapping user IDs.

\vspace{1mm}

\item Case~4~$\left( \casefour \right)$: In this case, each receiver gets its corresponding packet without any interference. We consider such packets to be delivered and no retransmission is required. Therefore, the transmitted packet of ${\sf Tx}_i$ leaves $Q_{i \rightarrow i}$ and joins $Q_{i,F}$, $i=1,2$.

\vspace{1mm}

\item Case~5~$\left( \casefive \right)$ and Case~6~$\left( \casesix \right)$: In these cases, ${\sf Rx}_1$ gets its corresponding packet interference free. We consider this packet to be delivered and no retransmission is required. Therefore, the transmitted packet of ${\sf Tx}_1$ leaves $Q_{1 \rightarrow 1}$ and joins $Q_{1,F}$, while the transmitted packet of ${\sf Tx}_2$ remains in $Q_{2 \rightarrow 2}$.

\vspace{1mm}

\item Case~7~$\left( \caseseven \right)$: In this case, ${\sf Rx}_1$ has a linear combination of the transmitted packets, while ${\sf Rx}_2$ has not received anything, see Table~\ref{table:16cases}. It is sufficient to provide the transmitted packet of ${\sf Tx}_2$ to both receivers. Therefore, the transmitted packet of ${\sf Tx}_2$ leaves $Q_{2 \rightarrow 2}$ and joins $Q_{2 \rightarrow \{ 1,2 \}}$. Note that the packet of ${\sf Tx}_1$ will not be retransmitted since upon delivery of the packet of ${\sf Tx}_2$, ${\sf Rx}_1$ can recover its corresponding packet. This packet leaves $Q_{1 \rightarrow 1}$ and joins $Q_{1,F}$. Similar argument holds for Case~8~$\left( \caseeight \right)$.

\vspace{1mm}

\item Cases~9,10,11, and 12: Similar to Cases~5,6,7, and 8 with swapping user IDs respectively.

\vspace{1mm}

\item Case~13~$\left( \nearrow \right)$: In this case, ${\sf Rx}_1$ has received the transmitted packet of ${\sf Tx}_2$ while ${\sf Rx}_2$ has not received anything, see Table~\ref{table:16cases}. Therefore, the transmitted packet of ${\sf Tx}_1$ remains in $Q_{1 \rightarrow 1}$, while the transmitted packet of ${\sf Tx}_2$ is required by ${\sf Rx}_2$ and it is available at ${\sf Rx}_1$. Hence, the transmitted packet of ${\sf Tx}_2$ leaves $Q_{2 \rightarrow 2}$ and joins $Q_{2 \rightarrow 2|1}$. Queue $Q_{2 \rightarrow 2|1}$ represents the packets at ${\sf Tx}_2$ that are available at ${\sf Rx}_1$, but ${\sf Rx}_2$ needs them.

\vspace{1mm}

\item Case~14~$\left( \searrow \right)$: This is similar to Case~13.

\vspace{1mm}

\item Case~15~$\left( \casefifteen \right)$: In this case, ${\sf Rx}_1$ has received the transmitted packet of ${\sf Tx}_2$ while ${\sf Rx}_2$ has received the transmitted packet of ${\sf Tx}_1$, see Table~\ref{table:16cases}. In other words, the transmitted packet of ${\sf Tx}_2$ is available at ${\sf Rx}_1$ and is required by ${\sf Rx}_2$; while the transmitted packet of ${\sf Tx}_1$ is available at ${\sf Rx}_2$ and is required by ${\sf Rx}_1$. Therefore, we have transition from $Q_{i \rightarrow i}$ to $Q_{i \rightarrow i|\bar{i}}$, $i=1,2$.

\vspace{1mm}

\item Case~16: Packet of ${\sf Tx}_i$ remains in $Q_{i \rightarrow i}$, $i=1,2$.

\end{itemize}

Phase~$1$ goes on for $\left( 4/3 \right) m + m^{\frac{2}{3}}$
%\begin{align}
%\left( 4/3 \right) m + m^{\frac{2}{3}}
%\end{align}
time instants, and if at the end of this phase, either of the queues $Q_{i \rightarrow i}$ is not empty, we declare error type-\cnt and halt the transmission (we assume $m$ is chosen such that $m^{\frac{2}{3}} \in \mathbb{Z}$).

Assuming that the transmission is not halted, let $N_{i,{\sf C}_1}$, $N_{i \rightarrow j|\bar{j}}$, and $N_{i \rightarrow \{ 1,2\}}$ denote the number of packets in queues $Q_{i,{\sf C}_1}$, $Q_{i \rightarrow j|\bar{j}}$, and $Q_{i \rightarrow \{ 1,2\}}$ respectively at the end of the transitions, $i=1,2$, and $j = i,\bar{i}$. The transmission strategy will be halted and an error type-\cnt will occur if any of the following events happens.
\begin{align}
\label{eq:errortypeI}
& N_{i,{\sf C}_1} > \mathbb{E}[N_{i,{\sf C}_1}] + m^{\frac{2}{3}} \overset{\triangle}= n_{i,{\sf C}_1},~ i=1,2; \nonumber \\
& N_{i \rightarrow j|\bar{j}} > \mathbb{E}[N_{i \rightarrow j|\bar{j}}] + m^{\frac{2}{3}} \overset{\triangle}= n_{i \rightarrow j|\bar{j}}, ~i=1,2, \text{~and~} j = i,\bar{i}; \nonumber \\
& N_{i \rightarrow \{ 1,2 \}} > \mathbb{E}[N_{i \rightarrow \{ 1,2 \}}] + m^{\frac{2}{3}} \overset{\triangle}= n_{i \rightarrow \{ 1,2 \}},~  i=1,2.
\end{align}

From basic probability, we know that
\begin{align}
\label{eq:expectedvalues}
& \mathbb{E}[N_{i,{\sf C}_1}] = \frac{\Pr\left( \mathrm{Case~1} \right) m}{1 - \sum_{i=9,10,13,16}{\Pr\left( \mathrm{Case~i} \right)}}  = m/12,  \\
& \mathbb{E}[N_{i \rightarrow i|\bar{i}}] = \mathbb{E}[N_{i \rightarrow \{ 1,2 \}}] = m/6,~\mathbb{E}[N_{i \rightarrow \bar{i}|i}] = m/12. \nonumber
\end{align}

%& \mathbb{E}[N_{i \rightarrow i|\bar{i}}] = \frac{\sum_{j=14,15}{\Pr\left( \mathrm{Case~j} \right)}}{1 - \sum_{i=9,10,13,16}{\Pr\left( \mathrm{Case~i} \right)}} m = \left( 1/6 \right) m, \nonumber \\
%& \mathbb{E}[N_{i \rightarrow \bar{i}|i}] = \frac{\Pr\left( \mathrm{Case~2} \right)}{1 - \sum_{i=9,10,13,16}{\Pr\left( \mathrm{Case~i} \right)}} m = \left( 1/12 \right) m, \nonumber \\
%& \mathbb{E}[N_{i \rightarrow \{ 1,2 \}}] = \frac{\sum_{j=11,12}{\Pr\left( \mathrm{Case~j} \right)}}{1 - \sum_{i=9,10,13,16}{\Pr\left( \mathrm{Case~i} \right)}} m = \left( 1/6 \right) m.

Furthermore, we can show that the probability of errors of types I and II decreases exponentially with $m$. More precisely, we use Chernoff-Hoeffding bound\footnote{We consider a specific form of the Chernoff-Hoeffding bound~\cite{Hoeffding} described in~\cite{Chernoff}, which is simpler to use and is as follows. If $x_1,\ldots,x_r$ are $r$ independent random variables, and $w=\sum_{i=1}^r{x_i}$, then $Pr\left[ |w-\mathbb{E}\left[ w \right] | > \alpha \right] \leq 2 \exp \left( \frac{-\alpha^2}{4 \sum_{i=1}^r \mathrm{Var} \left( x_i \right)} \right)$.}, to bound the error probabilities of types I and II. For instance, we have
\begin{align}
\Pr & \left[ \mathrm{error~type \noindent - \noindent I} \right] \leq  \sum_{i=1}^2{\Pr \left[ Q_{i \rightarrow i} \mathrm{~is~not~empty} \right]} \nonumber \\
& \leq 4 \exp \left( -m^{4/3}/(m+(3/4)m^{2/3}) \right),
% & = 4 \exp \left( -m^{4/3}/ \left( 3/4 \left[ \left( 20/9 \right) m + m^{\frac{2}{3}} \right] \right) \right),
\end{align}
which decreases exponentially to zero as $m \rightarrow \infty$.

At the end of Phase $1$, we add $0$'s (if necessary) in order to make queues $Q_{i,{\sf C}_1}$, $Q_{i \rightarrow j|\bar{j}}$, and $Q_{i \rightarrow \{ 1,2\}}$ of size equal to $n_{i,{\sf C}_1}$, $n_{i \rightarrow j|\bar{j}}$, and $n_{i \rightarrow \{ 1,2\}}$ respectively as defined in (\ref{eq:errortypeI}), $i=1,2$, and $j = i,\bar{i}$. For the rest of this section, we assume that Phase 1 is completed and no error has occurred.

We now use the ideas described in Section~\ref{Sec:Opportunities}, to further create packets of common interest. In particular, we demonstrate how to incorporate the ideas of Section~\ref{Sec:Opportunities} to create packets of common interest in an optimal way.
\begin{itemize}

\item {\bf Type I} Combining the packets in $Q_{i,{\sf C}_1}$ and $Q_{i \rightarrow i|\bar{i}}$ ({\it packet delivery with side information}): Consider the packets that were transmitted in Cases $1$ and $15$, see Fig.~\ref{Fig:CodingTypeI}. Providing $\left( a_1 + a_2 \right)$ and $\left( b_1 + b_2 \right)$ to both receivers is sufficient to recover the packets. Hence, we can remove two packets in $Q_{i,{\sf C}_1}$ and $Q_{i \rightarrow i|\bar{i}}$, by inserting their summation in $Q_{i \rightarrow \{ 1,2 \}}$, $i=1,2$, and then deliver this packet of common interest to both receivers during the second phase.

%\begin{figure}[ht]
%\centering
%\includegraphics[height = 2.75 cm]{Figures/case1-15.pdf}
%\caption{Suppose at a time instant, transmitters one and two send out packets $a_1$ and $b_1$ respectively, and later using Delayed-CSIT, transmitters figure out Case 1 occurred at that time. At another time instant, suppose transmitters one and two send out packets $a_2$ and $b_2$ respectively, and later using Delayed-CSIT, transmitters figure out Case 15 occurred at that time. Now, $\left( a_1 + a_2 \right)$ available at ${\sf Tx}_1$ and $\left( b_1 + b_2 \right)$ available at ${\sf Tx}_2$ are useful for both receivers and they are packets of common interest. Therefore, $\left( a_1 + a_2 \right)$ and $\left( b_1 + b_2 \right)$ can join $Q_{1 \rightarrow \{ 1,2 \}}$ and $Q_{2 \rightarrow \{ 1,2 \}}$ respectively.\label{fig:1and15}}
%\end{figure}
 
We have $\mathbb{E}[N_{i,{\sf C}_1}] < \mathbb{E}[N_{i \rightarrow i|\bar{i}}]$. Therefore, after this combination, queue $Q_{i,{\sf C}_1}$ becomes empty and we have 
\begin{align}
\mathbb{E}[N_{i \rightarrow i|\bar{i}}] - \mathbb{E}[N_{i,{\sf C}_1}] =  1/12 m
\end{align}
packets left in $Q_{i \rightarrow i|\bar{i}}$.

\item {\bf Type II} Combining packets in $Q_{i \rightarrow \bar{i}|i}$ and $Q_{i \rightarrow i|\bar{i}}$ ({\it interference delivery with side information}): Consider the packets that were transmitted in Cases $2$ and $14$, see Fig.~\ref{Fig:2and14}. If we provide $\left( a_1 + a_2 \right)$ to \emph{both} receivers then ${\sf Rx}_1$ can recover packets $a_1$ and $a_2$, whereas ${\sf Rx}_2$ can recover packet $b_1$. Therefore, $\left( a_1 + a_2 \right)$ is a packet of common interest and can join $Q_{1 \rightarrow \{ 1,2 \}}$. Hence, we can remove two packets in $Q_{1 \rightarrow 2|1}$ and $Q_{1 \rightarrow 1|2}$, by inserting their summation in $Q_{1 \rightarrow \{ 1,2 \}}$, and we deliver this packet of common interest to both receivers during the second phase. Note that due to the symmetry of the channel, similar argument holds for $Q_{2 \rightarrow 1|2}$ and $Q_{2 \rightarrow 2|1}$.

\begin{figure}[ht]
\vspace{-2.5mm}
\centering
\includegraphics[height = 2.75 cm]{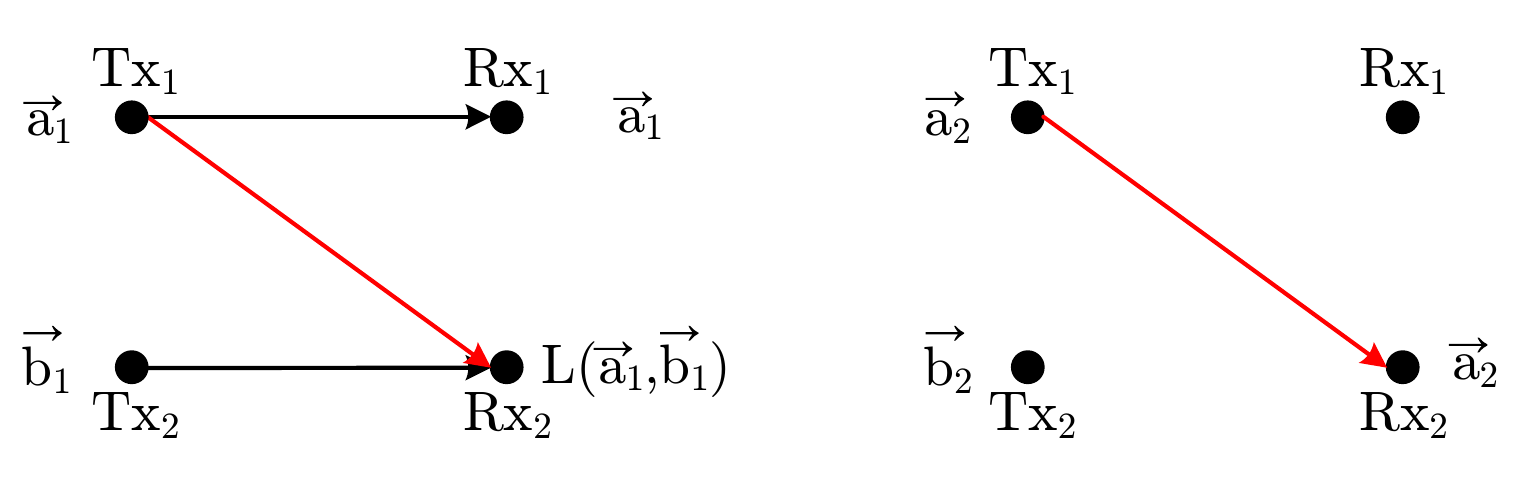}
\caption{Suppose at a time instant, transmitters $1$ and $2$ send out packets $a_1$ and $b_1$ respectively, and Case~2 occurred. At another time instant, suppose transmitters $1$ and $2$ send out packets $a_2$ and $b_2$ respectively, and Case~14 occurred. Now, $\left( a_1 + a_2 \right)$ available at ${\sf Tx}_1$ is useful for both receivers and it is a packet of common interest. Hence, $\left( a_1 + a_2 \right)$ joins $Q_{1 \rightarrow \{ 1,2 \}}$.\label{Fig:2and14}}
\vspace{-2.5mm}
\end{figure}
  
After combining the packets, queue $Q_{i \rightarrow i|\bar{i}}$ and $Q_{i \rightarrow \bar{i}|i}$ both become empty, $i=1,2$.

\end{itemize}

Hence at the end of Phase $1$, if the transmission is not halted, we have a total of 
{\small \begin{align}
& (4/3) \left[ \underbrace{1/8 m + m^{2/3}}_{\mathrm{Cases~11~and~12}} + \underbrace{1/8 m + m^{2/3}}_{\mathrm{coding~opportunities}}  \right] = 1/3 m + 8/3 m^{2/3}
\end{align}}
number of packets in $Q_{1 \rightarrow \{1,2\}}$ (same for $Q_{2 \rightarrow \{1,2\}}$).

This completes the description of Phase $1$. We now describe how to deliver the packets of common interest in Phase~2 of the transmission strategy. The problem resembles a network with two transmitters and two receivers where each transmitter ${\sf Tx}_i$ wishes to communicate its $m$ packets to {\it both} receivers, $i=1,2$. The channel gain model is the same as described in Section~\ref{Sec:System}. We refer to this network as the two-multicast network, and we have the following result for it.

%\begin{figure}[ht]
%\centering
%\includegraphics[height = 3.5 cm]{Figures/two-multicast.pdf}
%\caption{Two-multicast network. Transmitter ${\sf Tx}_i$ wishes to reliably communicate message $\hbox{W}_i$ to both receivers, $i=1,2$. The capacity region with no, delayed, or instantaneous CSIT is the same.\label{fig:two-multicast}}
%\end{figure}

\begin{lemma}
\label{lemma:multicast}
For the two-multicast network as described above, the optimal throughput region is given by
\begin{equation}
\label{}
\left\{ \begin{array}{ll}
\vspace{1mm} R_i \leq 1/2, & i = 1,2,\\
R_1 + R_2 \leq 3/4. &
\end{array} \right.
\end{equation}
\end{lemma}

This result basically shows that the capacity region of the two-multicast network described above is equal to the capacity region of the multiple-access channel formed at either of the receivers. The proof of Lemma~\ref{lemma:multicast} is omitted due to space limitations and can be found in~\cite{vahid2013capacity} Section~V. Basically, transmitters can create enough random linear equations of their packets such that the receivers can recover all packets from these equations with probability $1$ as $m \rightarrow \infty$.

\noindent {\bf Phase 2} [transmitting packets of common interest]: In this phase, we deliver the packets in $Q_{1 \rightarrow \{ 1,2 \}}$ and $Q_{2 \rightarrow \{ 1,2 \}}$ using the transmission strategy for the two-multicast problem. More precisely, the packets in $Q_{i \rightarrow \{ 1,2 \}}$ will be considered as the packets of ${\sf Tx}_i$.From Lemma~\ref{lemma:multicast}, we know that rate tuple $\left( R_1, R_2 \right) = \left( 3/8, 3/8 \right)$ is achievable. Therefore, transmission of the packets in $Q_{1 \rightarrow \{ 1,2 \}}$ and $Q_{2 \rightarrow \{ 1,2 \}}$, will take
\begin{align} 
t_{\mathrm{total}} = \left( 2/3 m + 16/3 m^{2/3} \right) / (3/4) \raisebox{2pt}{.}
\end{align}

Therefore, the total transmission time of our two-phase achievability strategy is equal to
\begin{align}
4/3 m + m^{\frac{2}{3}} + \left( 2/3 m + 16/3 m^{2/3} \right) / (3/4) \raisebox{2pt}{,}
\end{align}
hence, if we let $m \rightarrow \infty$, the decoding error probability of recovering packets of common interest goes to zero, and we achieve a symmetric throughput of 
\begin{align}
R_1 = R_2 = \lim_{\substack{\epsilon,\delta \rightarrow 0 \\ m \rightarrow \infty}}{\frac{m}{t_{\mathrm{total}}}} = 0.45.
\end{align}

This completes the achievability proof of corner point $A$. Although, we have only provided the transmission strategy for $p = 1/2$, same concepts and ideas apply to the general value of $p$. Therefore, we achieve the following throughput region.
\begin{equation}
\label{eq:AchRegion}
\left\{ \begin{array}{ll}
\vspace{1mm} 0 \leq R_i \leq p, & i = 1,2, \\
R_i + (2-p) R_{\bar{i}} \leq p(2-p)^2, & i = 1,2,
\end{array} \right.
\end{equation}

%%%%%%%%%%%%%%%%%%%%%%%%%%%%%%%%%%%%%%%%%%%%%%%%%%%
\section{Proving the Optimality}
\label{Sec:Outerbound}

In this section, we prove the optimality of the transmission scheme we described in Section~\ref{Sec:Transmission}. In particular, we derive an outer-bound on the throughput region for the network with two transmitter-receiver pairs (introduced in Section~\ref{Sec:System}) that matches our achievable region in (\ref{eq:AchRegion}), thus, proving the optimality of our scheme. Suppose a throughput tuple of $\left( R_1,R_2 \right)$ is achievable, meaning that ${\sf Tx}_i$ has $n R_i$ packets for ${\sf Rx}_i$, and uses a beamforming precoding matrix $\mathbf{V}_i^n$ to create its transmit signal, $i=1,2$. The received signal at ${\sf Rx}_i$ is
\begin{align}
\vec{y}_i^n = \mathbf{G}_{1i}^n \mathbf{V}_1^n \mathbf{A} + \mathbf{G}_{2i}^n \mathbf{V}_2^n \mathbf{B}, \quad i = 1,2,
\end{align}
and the decodability condition is to have
\begin{align}
\label{eq:decodabilityend}
\mathrm{dim} & \left( \mathrm{Proj}_{\mathcal{I}_i^c}~\mathrm{colspan}\left( \mathbf{G}_{ii}^n \mathbf{V}_{i}^n \right) \right) \nonumber \\
& = \mathrm{dim}\left( \mathrm{colspan}\left( \mathbf{V}_{i}^n \right) \right) = n R_i,  \quad i = 1,2,
\end{align}
with probability $1$. We first derive the outer-bound on individual throughput rates. We have
\begin{align}
n & R_1 \overset{a.s.}= \mathbb{E}\left[ \mathrm{dim}\left( \mathrm{Proj}_{\mathcal{I}_1^c}~\mathrm{colspan}\left( \mathbf{G}_{11}^n \mathbf{V}_{1}^n \right) \right) \right] \nonumber \\
& \overset{(a)}\leq \mathbb{E}\left[ \mathrm{rank}\left[ \mathbf{G}_{11}^n \mathbf{V}_{1}^n \right] \right] \overset{(b)}\leq p n,
% \overset{(c)}\leq \mathbb{E}\left[ \mathrm{rank}\left[ \mathbf{G}_{11}^n \mathbf{V}_{1}^n \right] \right] + o(n) \nonumber \\
\end{align}
where the first equality holds since the decodability condition should be satisfied with probability $1$; $(a)$ is true since we ignored the interference, and this can only increase the dimension; and $b$ holds since $\alpha_{11}(t)$ is equal to $1$ with probability $p$. Dividing both sides by $n$, we get $R_1 \leq p$. Similarly, we get $R_2 \leq p$. Next, we derive the following outer-bound: 
\begin{align}
\label{eq:lastbound}
R_i + (2-p) R_{\bar{i}} \leq p(2-p)^2, \quad i = 1,2.
\end{align}
Again, by symmetry, we only prove the result for $i=1$. We use the following lemma in the proof of the outer-bound.

\begin{lemma}
\label{KeyLemma}
The following inequality holds almost surely under the Delayed-CSIT assumption (for $0 < p \leq 1$).
\begin{align}
\mathbb{E}\left[ \mathrm{rank}\left[ \mathbf{G}_{12}^n \mathbf{V}_{1}^n \right] \right] \ge \frac{1}{2-p} \mathbb{E}\left[ \mathrm{rank}\left[ \mathbf{G}_{11}^n \mathbf{V}_{1}^n \right] \right].
\end{align}
\end{lemma}
\vspace{-3.3mm}
{\it Proof:}
\begin{align}
\mathbb{E} & \left[ \mathrm{rank}\left[ \mathbf{G}_{12}^n \mathbf{V}_{1}^n \right] \right] \nonumber \\
& = \mathbb{E}\left[ \sum_{t=1}^n{\mathrm{rank}\left[ \mathbf{G}_{12}^t \mathbf{V}_{1}^t \right] - \mathrm{rank}\left[ \mathbf{G}_{12}^{t-1} \mathbf{V}_{1}^{t-1} \right]} \right] \nonumber \\
& = \mathbb{E}\left[ \sum_{t=1}^n{\mathbf{1}{\left\{ \alpha_{12}(t) g_{12}(t) \vec{v}_{1}(t) \notin \mathrm{rowspan}\left( \mathbf{G}_{12}^{t-1} \mathbf{V}_{1}^{t-1} \right) \right\} }} \right] \nonumber \\
& \overset{a.s.}= p \mathbb{E}\left[ \sum_{t=1}^n{\mathbf{1}{\left\{ \vec{v}_{1}(t)~\notin~\mathrm{rowspan}\left( \mathbf{G}_{12}^{t-1} \mathbf{V}_{1}^{t-1} \right) \right\} }} \right] \\
& \ge p \mathbb{E}\left[ \sum_{t=1}^n{\mathbf{1}{\left\{ \vec{v}_{1}(t)~\notin~\mathrm{rowspan}\begin{bmatrix} \mathbf{G}_{12}^{t-1} \mathbf{V}_{1}^{t-1} \\ \mathbf{G}_{11}^{t-1} \mathbf{V}_{1}^{t-1} \end{bmatrix}  \right\} }} \right] \nonumber \\
& \overset{(a)}= \frac{p}{p(2-p)} \mathbb{E}\left[ \sum_{t=1}^n{ \mathrm{rank} \begin{bmatrix} \mathbf{G}_{12}^t \mathbf{V}_{1}^t \\ \mathbf{G}_{11}^t \mathbf{V}_{1}^t \end{bmatrix} - \mathrm{rank} \begin{bmatrix} \mathbf{G}_{12}^{t-1} \mathbf{V}_{1}^{t-1} \\ \mathbf{G}_{11}^{t-1} \mathbf{V}_{1}^{t-1} \end{bmatrix} } \right] \nonumber \\
& = \frac{1}{2-p} \mathbb{E}\left[ \mathrm{rank} \begin{bmatrix} \mathbf{G}_{12}^n \mathbf{V}_{1}^n \\ \mathbf{G}_{11}^n \mathbf{V}_{1}^n \end{bmatrix}\right] \ge \frac{1}{2-p} \mathbb{E}\left[ \mathrm{rank}\left[ \mathbf{G}_{11}^n \mathbf{V}_{1}^n \right] \right], \nonumber 
\end{align}
where the third equality holds almost surely since $g_{12}(t)$ comes from a continuous distribution; and $(a)$ holds since
\begin{align}
\label{eq:bothoffprob} 
\vspace{-2.75mm}
\Pr \left[ \alpha_{12}(t) = \alpha_{11}(t) = 0 \right] = 1 -(1-p)^2 = p (2-p). \nonumber ~~\hfill{\blacksquare}
\end{align}
% \end{proof}

\vspace{-1.9mm}
To obtain the outer-bound in (\ref{eq:lastbound}) for $i=1$, we have
\begin{align}
& n \left( R_1 + (2-p) R_2 \right) \overset{a.s.}=  \mathbb{E}\left[ \mathrm{dim}\left( \mathrm{Proj}_{\mathcal{I}_1^c}~\mathrm{colspan}\left( \mathbf{G}_{11}^n \mathbf{V}_{1}^n \right) \right) \right] \nonumber \\
& + (2-p) \mathbb{E}\left[ \mathrm{dim}\left( \mathrm{Proj}_{\mathcal{I}_2^c}~\mathrm{colspan}\left( \mathbf{G}_{22}^n \mathbf{V}_{2}^n \right) \right) \right] \nonumber \\
& \overset{(a)}\leq \mathbb{E}\left[ \mathrm{rank}\left[ \mathbf{G}_{11}^n \mathbf{V}_{1}^n \right] \right] + (2-p) \mathbb{E}\left[ \mathrm{rank}\left[ \mathbf{G}_{12}^n \mathbf{V}_{1}^n + \mathbf{G}_{22}^n \mathbf{V}_{2}^n \right] \right] \nonumber \\
& - (2-p) \mathbb{E}\left[ \mathrm{rank}\left[ \mathbf{G}_{12}^n \mathbf{V}_{1}^n \right] \right] \\
& \overset{\mathrm{Lemma~\ref{KeyLemma}}}\leq (2-p) \mathbb{E}\left[ \mathrm{rank}\left[ \mathbf{G}_{12}^n \mathbf{V}_{1}^n + \mathbf{G}_{22}^n \mathbf{V}_{2}^n \right] \right] \overset{(c)}\leq p (2-p)^2 n, \nonumber
\end{align}
where the first equality holds since the decodability condition should be satisfied with probability $1$; $(b)$ is true since we ignored interference; and $(c)$ holds since
\begin{align}
\label{eq:bothoffprob}
%\vspace{-1mm}
\Pr \left[ \alpha_{12}(t) = \alpha_{22}(t) = 0 \right] = 1 -(1-p)^2 = p (2-p).
\end{align}
\vspace{-1mm}
Dividing both sides by $n$, we get the desired outer-bound. This completes the derivation of the outer-bound.

% \vspace{-1mm}

\section{Conclusion and Future Directions}
\label{Sec:Conclusion}

In this work, we considered a physical layer model for packet networks that allows for the flexibility of storing analog signals at the receivers (when collision occurs), and utilizing them for decoding packets in the future. We proposed two general coding opportunities at the transmitters for interference management and a unified transmission strategy that systematically utilizes such opportunities in the context of a network with two transmitter-receiver pairs. We further prove the optimality of our scheme by developing an outer-bound via an extremal rank inequality with Delayed-CSIT assumption.

An extension of our work would be to consider scenarios in which more than two transmitter-receiver pairs can coordinate for interference management. However, obtaining global channel state information (even delayed) in such cases is quite difficult. Thus, it would be interesting to study a larger network in which transmitters have access to local~\cite{aggarwal2011achieving,vahid2010capacity} and/or mismatched~\cite{reddy2012distributed} channel state information, and see how they can coordinate based on such knowledge to perform interference management. 
%An extension of our work would be to consider larger wireless packet networks. However, obtaining global channel state information in larger networks is practically infeasible.
% In this work, we demonstrated how stored analog signals at the receivers can be utilized to improve the achievable throughput region. In particular, we considered a new physical layer model, and we proposed two novel coding opportunities 

%\vspace{-2mm}
%%%%%%%%%%%%%%%%%%%%%%%%%%%%%%%%%%%%%%%%%%%%%%%%%%%
\bibliographystyle{ieeetr}
\bibliography{bib_infocom}

\end{document}